# Bar Ilan University

## Department of Chemistry

## M.Sc. Thesis

**Combinatorial approach for development of new metal oxides materials for all oxide photovoltaics.**

**Advisor:**

Prof. Arie Zaban

**By:**

Klimentiy Shimanovich     324520303

Ramat Gan, Israel                                        2014

This work was carried out under the supervision of

Prof. Arie Zaban

Department of Chemistry, Bar-Ilan University

## Chapter I: Introduction

## Chapter II: Research aims

## Chapter III: Experimental Section

## Chapter IV: Characterization methods

## Chapter V: Results and discussion

## Chapter VI: Summary and Conclusions

## Chapter VIII: Bibliography

# List of Figures

## List of Tables

# Abstract

The combinatorial approach to all oxide material and device research is based on the synthesis of hundreds of related materials in a single experiment. This approach requires the development of new tools to rapidly characterize these materials libraries and new techniques to analyze the resulting data. The research presented here is intended to make a contribution towards meeting this demand, and thereby advance the pace of materials research. In many cases photovoltaic determinations are well-suited for high throughput methodologies, enabling direct quantitative analysis of properties whose implementation I demonstrate in the first part of my thesis.

The first part of this thesis discusses the development of a material characterization tool called the "Electrical conductivity scanning system", which is utilized to determine resistivity, conductivity and activation energy for compositionally graded metal oxide libraries.

The second part of this thesis discusses the spatial solution of optical and electrical properties for an inhomogeneous NiO thin film prepared by pulsed laser deposition. The structural, electrical and optical properties of NiO, as well as a detailed characterization of activation energy and its correlation with texture factor, are presented. It was found that the NiO thin film fabricated by PLD can be used as a hole-selective contact in multilayer thin film All-Oxide photovoltaic cells and can potentially improve the photovoltaic performance due to its selective free charge carrier transport and ambient stability.

The third part of my thesis is focused on a combination of the knowledge obtained in development of high throughput analytical tools (which are discussed in the first part) and a selection of the metal oxide material for improved PV performance (discussed in the second part) to create combinatorial multi metal oxide photovoltaic cells.

This thesis focuses on the development and utilization of high throughput and combinatorial methods that have incorporated, or are associated with, the all-oxide photovoltaic field. The development of new absorbers often requires novel buffer layers, contact materials, and interface engineering. The importance and contribution of the combinatorial material science approach for the development and rapid characterization of new and known metal oxide thin films for all oxide photovoltaics have been shown and discussed in my thesis.

# Chapter I: Introduction

## 1.1 Introduction

Many technological solutions depend on the development of new materials with improved properties, and increasingly on materials that can be optimized to perform more than one function. Combinatorial and High Throughput materials science is an approach used for the rapid discovery, study, and optimization of new and known materials that combines rapid synthesis, high throughput testing, and high-capacity information processing to prepare, analyze, and interpret large numbers of diverse material compositions.[1,2]

The word combinatorial refers to the systematic variation of parameters such as composition, thickness, temperature, or other single variables to explore a wide parameter space that is determined by a given high throughput experiment.

High throughput experiments in materials research are defined by their types of materials, and typically require the preparation of an array of materials' libraries. In addition, it requires a fast method to screen materials' properties and suitable software for experimental control, data storage, data analysis, and experimental design. These experiments are generally achieved through the use of rapid serial automation measurements and typically have a large combination of material variables.[3]

A high throughput experiment often starts with a first set of experiments that cover the parameter space selected. The most effective materials are determined during the early stage of the study and are usually used for the all sets of later experiments. Thus, the material with the most desirable properties became the object of the research focus.

Many useful discoveries in the field of materials science have been made as a result of adopting combinatorial approaches, particularly in the areas of optics[4], dielectrics[5], superconductivity[6] and magnetic materials.[7]

More commonly cited examples[4, 5, 6, 7] include the development of new generation capacitance materials for random access memory devices. The traditional capacitor, widely used in solar batteries, is amorphous silica. However, amorphous silica is being replaced with materials based on $ZrO_2$-$SnO_2$-$TiO_2$ for better performance.[8]

In our current work, we focus on the development of new metal oxides that potentially can improve the performance of solar cells. Metal oxides' abundance, non-toxicity and long-term stability allows for the possibility of constructing durable, long-lasting solar cells with metal oxides as the active light-absorbing component.

The second half of the research focuses on the design, construction and development of scanner-based analytical devices for metal oxide coating characterization. The scanner is identified as a High Throughput Combinatorial analytical and characterization method in the field of conducting and photoactive thin film metal oxide materials. The general strategy for development and construction of combinatorial and High Throughput materials is shown in Figure 1.

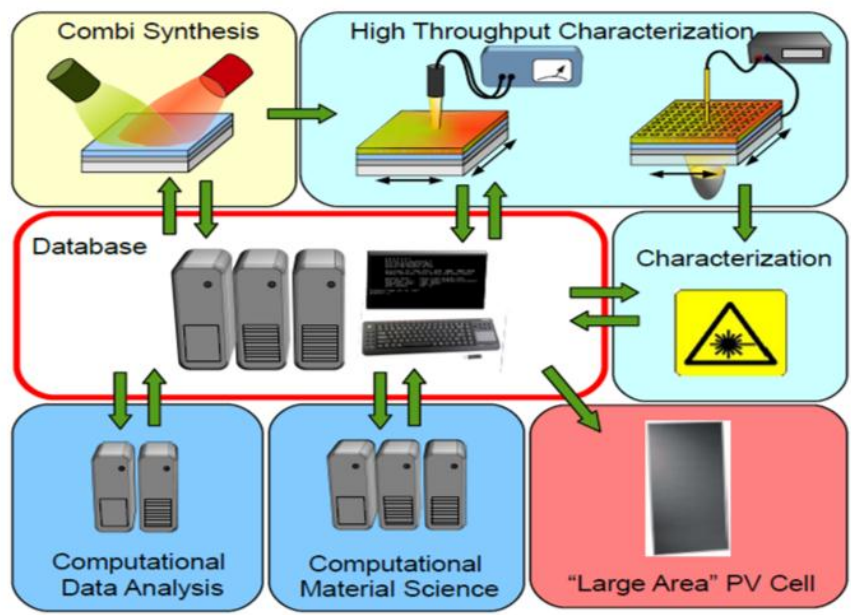

**Figure 1.** Schematic representation of Combinatorial and High Throughput materials development.

## 1.2 Solar energy

The world's total energy consumption for 2010 was estimated at about $2*10^{14}$ KWh, of which about 1/10 was used as electricity[9]. Electricity is the basis of all civilized countries and industrialization, and its access is a fundamental step towards achieving people's welfare. About 65% of the total electricity consumption is obtained from fossil fuels (coal, oil and gas) and only 20% comes from nuclear power plants and the rest is produced by renewable sources, such as hydropower, biomasses, **PV** and geothermal sources. These renewable sources are seen as the most promising way of granting electricity to the whole world, and can help reach a self-sustaining energy system. There is a striking need for alternative sources of energy such as solar energy. However, the performance efficiency of existing solar cells needs to be improved, and there is now a lot of interest in research that can discover new types of materials for solar applications.

It has been calculated that the total amount of solar energy reaching the Earth's surface is more than 10,000 times the world's total energy consumption. It is obvious that this source can supply the substantial part or even all of our future energetic needs. The solar power does not necessarily require a grid connection, is noiseless, maintenance free, reliable for more than 20 years and can be integrated in consumer electronics for low-power applications.

## 1.3 Solar cells

Solar cells can be fabricated from a number of semiconductor materials, most commonly silicon (Si) – crystalline, polycrystalline, and amorphous. It can also be fabricated from other semiconductor materials such as GaAs, GaInP, Cu(InGa)Se$_2$, and CdTe. Solar cell materials, in addition to low-cost fabrication, are chosen largely on the basis of high electronic stability, conduction nature, carrier's generation and recombination, and absorption characteristics which match the solar spectrum.[10]

Photovoltaic energy conversion in solar cells consists of two general steps. First, absorption of light generates an electron-hole pair. Second, separation of the electron and hole by the structure of the device: electrons to the negative terminal and holes to the positive terminal, thus generating electrical power. There are 5 main types of solar cells: 1) Amorphous silicon solar cells, 2) Crystalline and Polycrystalline silicon solar cells, 3) Thin film solar cells, 4) Dye-sensitized solar cells and Quantum dot sensitized solar cells, 5) Organic solar cells. Aeneral description of typical solar cells is presented below.

**Amorphous silicon solar cells**[11,12]**:** Hydrogenated amorphous silicon (a-Si:H) and silicon germanium (a-SiGe:H), are two major intrinsic layers used in this type of solar cell. The key technique of making a-Si:H useful in electronic devices is hydrogenation, where hydrogen atoms terminate silicon dangling bonds and reduce the defect density. The largest thickness for such amorphous silicon solar cells is related to a-Si:H/a-SiGe:H/a-SiGe:H triple-junction solar cells and is about half a micrometer. Such low thickness is required due to the high absorption and high generation rate of free charge carriers of amorphous silicon solar cells. The highest stable cell efficiency reported for this type of cell is 13.0%.

**Crystalline and Polycrystalline silicon solar cells**[11]**:** Two types of crystalline solar cells exist: 1) Mono-crystalline silicon, produced by slicing wafers from high purity crystal ingot, 2) Multi crystalline silicon, made by sawing a cast block of silicon first into bars and then into wafers. Mono crystalline silicon solar cells have higher efficiencies than multi crystalline silicon solar cells. In crystalline silicon photovoltaics, solar cells are generally connected and then laminated under toughened, high transmittance glass to produce weather resistant photovoltaic modules.

**Thin film solar cells**[9,13] – are based on thin layers of material of 1-3 micrometers thickness that strongly absorb light. The electrons freed by the photons travel only a short distance inside the cell to the cell contacts and from there to an external circuit to produce power. This type of behavior reduces the demand for high purification and crystallinity of the material, eliminating one of the causes for the high cost of solar cells. Thin film solar cells typically consist of 5-10 different layers whose functions include reducing resistance, forming the *p-n* junction, reducing reflection losses, and providing a robust layer for contact and interconnection between cells. These cells are typically made from $Cu(In,Ga)Se_2$, CdTe, and

a-Si:H materials. The maximum efficiency reported in 2013 by NREL was for solar cells made from CIGS[12] – 20.4%.

**Dye and quantum dot sensitized solar cells**[14] – In DSSC and QDSSC light is absorbed by a sensitizer (either dye or quantum dots), which is anchored to the surface of a wide band gap semiconductor. Charge separation takes place at the interface via photo-induced electron injection from the sensitizer into the conduction band of the semiconductor. Carriers are transported in the conduction band of the semiconductor to the charge collector. Efficiencies for these cells reach, in DSSC[12] 11.5%, and in QDSSC[15] 6.5%.

**Organic solar cells**[16] – are economically viable for large scale power generation. Organic small-molecule and polymer materials are inherently inexpensive. This type of cell has very high optical absorption coefficients in relatively narrow absorption spectra that permit the use of films with thicknesses of only several hundred nanometers. The organic solar cells are compatible with plastic substrates and fabricated using high throughput, low-temperature approaches that employ one of a variety of well-established printing techniques in a roll-to-roll process. The highest reported efficiency in 2013 by NREL was of Tandem organic solar cells – 12%.

## 1.4 Current obstacles in solar cells

The well-known Si solar cells suffer from high production cost and difficulties in material purification methods. The major complication in Si solar cells is reduction in power conversion efficiencies in terms of high temperature, oxidation with time, and hazardous (toxic) waste in its production. Moreover, solar cells based on CIGS and CdTe have a tendency to degrade with an increase in temperature and humidity. The encapsulation of solar cells effect degradation modes. Devices made of amorphous silicon thin films suffer from losses of 50% or more in power output over the first hundreds of hours of performance. This loss leads to Si-H bonds degradation due to light radiation which decreases efficiency.

## 1.5 Methods for improvement solar cells

Solar cell instability can be minimized by manipulating layer thicknesses and device structures. Moreover, simultaneous achievement of the following goals results in the creation of a new type of efficient solar cell based on metal oxides:

**1)** Doping or mixing relatively narrow band absorbers based on CuO, $Cu_2O$, $Co_3O_4$, CoO, metal oxides with wide band gap metal oxides such as: NiO, ZnO, $V_2O_5$, $In_2O_3$, $WO_3$ or with metals (particularly Al and Li) can improve solar cell performance characteristics by minimization of optical losses such as reflectance and absorption, minimizing serial and shunt resistances, improving stability, and even increasing power conversion in terms of high temperatures etc.

**2)** Variation of thicknesses and types of metal contacts.

**3)** Fabrication of electron/hole selective contacts with desirable electrical and optical properties.

# Chapter II: Research aims

The following research aims are at the heart of the current research work: Analysis and electrical characterization of new types of thin film metal oxides using a high throughput conductivity scanner for multilayer thin film all oxide photovoltaic cells, which requires:

a) Development of conductivity scanner as a High Throughput tool for electrical property characterization of new and known inhomogeneous metal oxide coatings.

b) Characterization of CuO and NiO thin film coatings for electron/hole generation and selective hole conduction respectively. Those materials were selected due to their cost effective properties that can improve photovoltaic cells made of metal oxides.

c) Investigation of CuO-NiO-In$_2$O$_3$ thin film metal oxide compositions as new potential absorbers for solar cells. The three metal components are supposed to provide an improvement in power conversion efficiencies of solar cells based on metal oxides.

# Chapter III: Experimental Section

## 3.1 Design and fabrication of combinatorial libraries for electrical analysis

For solar cell research purposes, first libraries of single, binary and ternary CuO, NiO and $In_2O_3$ thin film material compositions deposited onto glass substrates were generated for electrical, optical, structural and morphological properties analysis. The solar cells based on those compositions were fabricated and photovoltaic performance efficiencies were evaluated.

## 3.2 Pulsed laser deposition of CuO, NiO, Cu-Ni-O, CuO-NiO-$In_2O_3$ thin films

Pulsed Laser Deposition (PLD) is an ideal technique to be used for combinatorial approaches[17,18,19]. By simply changing the deposition targets we can obtain alternating layers with different periodicities both vertically and laterally, along the substrate surface. By changing the laser impact area location and the number of pulses on each target used for ablation, we can grow films with a continuous variation of chemical composition, which will be a function of the location on the substrate.

In current research CuO, NiO, $In_2O_3$ targets were purchased from Kurt J. Lesker Company and used for pulsed laser deposition. The CuO, NiO, Cu-Ni-O and CuO-NiO-$In_2O_3$ thin films were deposited using a pulsed laser deposition system (Coherent CompexPro102 Model, Neocera, US manufacture). The PLD system consists of a KrF excimer laser with an emission wavelength of 248nm and maximum pulse energy of 400 mJ, a target carrousel, a substrate heater up to 800 ºC, a linear translation stage for beam scanning and a sample holder with an adapter to accommodate square glass substrates with a side length of 71.3 mm. The schematic structure of the pulsed laser deposition apparatus is presented in Figure 2.

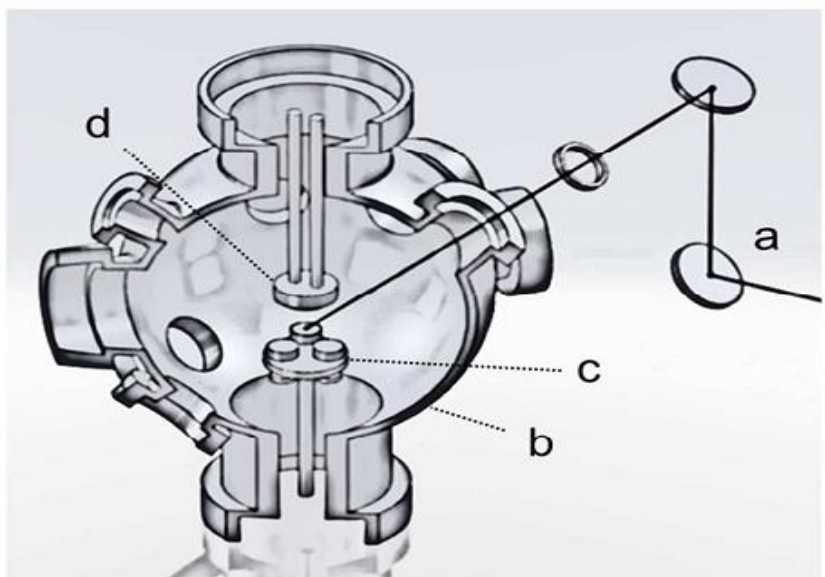

**Figure 2.** Schematic structure of Pulsed Laser Deposition setup: a) 248 nm KrF excimer laser. b) Vacuum chamber. c) Target on the target carrousel. d) Substrate holder.

### Fabrication of CuO and NiO thin films:

CuO absorber and NiO hole conductor thin films, with an inhomogeneous thickness profile, were produced by the pulsed laser deposition (PLD) technique onto glass substrates supplied by Hartford (thickness of 2.2 mm) and alkaline free glass substrates AF- 45 (Präzisions Glas & Optik, thickness of 1.1mm), with both substrates size of 71.2 x 71.2 $mm^2$. For pulsed laser deposition of CuO, the KrF laser operated at 8 Hz pulse rate, 97 mJ/pulse applied energy and ~1.54 $J/cm^2$ energy density. For NiO the deposition parameters were optimized using a pre-screening method and further applied to fabricate CuO thin films. The NiO deposition parameters were as follows: 8 Hz pulse rate, 106.1 mJ/pulse applied energy, ~1.7 $J/cm^2$ energy density. For deposition of CuO and NiO the following parameters were kept identical i.e.: measured active area of 6.3 mm$^2$, 30000 pulses, target substrate distance of 77mm with a constant oxygen pressure of ~130mTorr and temperature of 400°C. Both substrates were held in a fixed position in order to create CuO and NiO thin films with gradient thicknesses.

**Pre-screening process for NiO thin film:**

As a first step for material characterization and identification, we screened fourteen NiO samples (see Figure 3 and Table 1). Two PLD deposition parameters which are: 1) oxygen partial pressure and 2) distance between target and substrate, were identified as main parameters that affected the structural, optical and electrical properties of polycrystalline NiO films. Therefore, we focused on the sample library deposited at a relatively high oxygen pressure (~130mTorr) and high substrate to target distance (77mm).

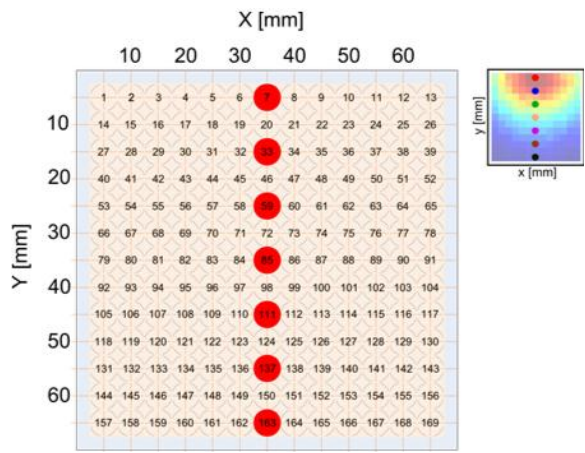

**Figure 3.** Process of pre-screening materials' properties through the maximal film thickness changes (middle column of seven points) to find optimal properties.

|   | Target substrate distance (mm) | Pressure (Torr) | Temperature (°C) | Laser energy (mJ) | Pulses |
|---|---|---|---|---|---|
| 1 | 47 | 0.00277 | 600 | 146.6 | 15000 |
| 2 | 47 | 0.00227 | 600 | 108.5 | 15000 |

| 3 | 77 | 0.134 | 400 | 109.2 | 30000 |
| 4 | 77 | 0.127 | 600 | 108.8 | 30000 |
| 5 | 77 | 0.0765 | 23 | 230 | 30000 |
| 6 | 77 | 0.00296 | 23 | 110.3 | 60000 |
| 7 | 77 | 0.00294 | 400 | 111.3 | 60000 |
| 8 | 77 | 0.00297 | 600 | 110 | 60000 |
| 9 | 77 | 0.13 | 400 | 109.6 | 30000 |
| 10 | 77 | 0.125 | 23 | 110.7 | 15000 |
| 11 | 77 | 0.133 | 400 | 110.3 | 30000 |
| 12 | 77 | 0.132 | 400 | 111 | 30000 |
| 13 | 77 | 0.132 | 400 | 110.6 | 30000 |
| 14 | 77 | 0.0131 | 600 | 110.2 | 30000 |

**Table1.** PLD deposition parameters of NiO coatings on glass substrates.

The bell shape profile of the film was created due to plasma differences. Therefore, the pre-screening process started at the center of the top edge of the substrate (point 7) where the plasma energy during deposition is high and ended at the center of the bottom edge of the sample (point 163 in Figure 3) where the deposition rate is low due to low plasma energy.

**Fabrication of Cu-Ni-O and CuO-NiO-In$_2$O$_3$ thin films:**

Cu-Ni-O and CuO-NiO-In$_2$O$_3$ thin film absorbers were deposited by using pulsed laser deposition (PLD) onto glass substrates which were supplied by Hartford (thickness of 2.2 mm). Depositions were carried out with target to sample distance of 77 mm, heater temperature of 400 °C, and an oxygen pressure of 130 mTorr. The effect of the laser inside the vacuum chamber with 104 mJ (energy) and a laser spot size of 6.3 mm$^2$ was measured and corresponded to an energy density of ~1.6 J/cm$^2$. In order to achieve the inhomogeneous thickness profile for binary Cu-Ni-O and ternary CuO-NiO-In$_2$O$_3$ compositions spread libraries, with varying composition from 0 to 100% along three different directions, we rotated the substrate by 120° each time and then the targets were ablated sequentially with 20 laser pulses for CuO, 10 laser pulses for NiO and 10 laser pulses for In$_2$O$_3$ with a repetition rate of 8Hz (see Figure 4). The ablation was repeated 1500 times until the total Cu-Ni-O and CuO-NiO-In2O3 film thickness reached the estimated value.

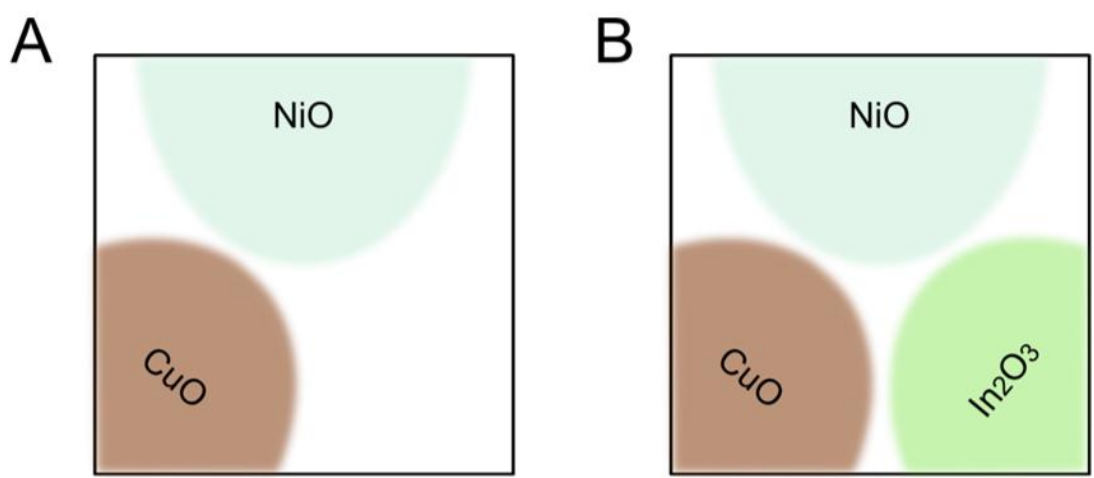

**Figure 4.** The complete libraries of: a) binary Cu-Ni-O materials system and b) ternary CuO-NiO-In2O3 materials system, deposited by PLD. The angles between each of the materials are 120°, which allowed for the presentation of the material compositions using a triangle phase diagram.

**Spray Pyrolysis of TiO$_2$ thin films:**

Titanium oxide is one of the most widely used thin film materials prepared by spray pyrolysis and used in solar cells. Hence, compact TiO$_2$ layers were deposited by spray pyrolysis onto commercially available fluorine doped SnO$_2$ (FTO) coated glass substrates with a size of 71.2 x 71.2 mm$^2$ and sheet resistance of 15 Ω per square (TEC 15, Hartford Glass Co.Inc.). The substrates were thoroughly washed with soap, rinsed with ethanol followed by de-ionized water, and dried under a dry air stream, then cleaned again by plasma and placed onto a Ceran hotplate (Harry Gestigkeit GmbH). A precursor solution of 0.1 M titanium tetraisopropoxide and 0.1 M acetylacetone in ethanol and isopropanol (mixing ratio 1: 1) was sprayed using an ultrasonic spray nozzle (Spraying Systems Co.) onto the substrates at a hotplate temperature of 450 °C. The alcoholic solvents were used due to their low surface tension and viscosity that facilitates the formation of small spray droplets while its low boiling point enables it to be efficiently removed from the deposition substrates in the vapor phase. The nozzle was mounted onto a commercial *x-y-z* scanner (EAS GmbH), the precursor flow rate of 60 $cm^3 h^{-1}$ was controlled using a syringe pump (Razel Scientific Instruments), while clean dehumidified compressed air, at a flow rate of 8 l $min^{-1}$, was used as a carrier gas. The *x-y* scan velocity was 30 mm $s^{-1}$, and the nozzle to substrate distance was approximately 6.9 cm. For combinatorial device fabrication a linear thickness gradient was produced using a series of spray cycles with a successively decreasing scan area.

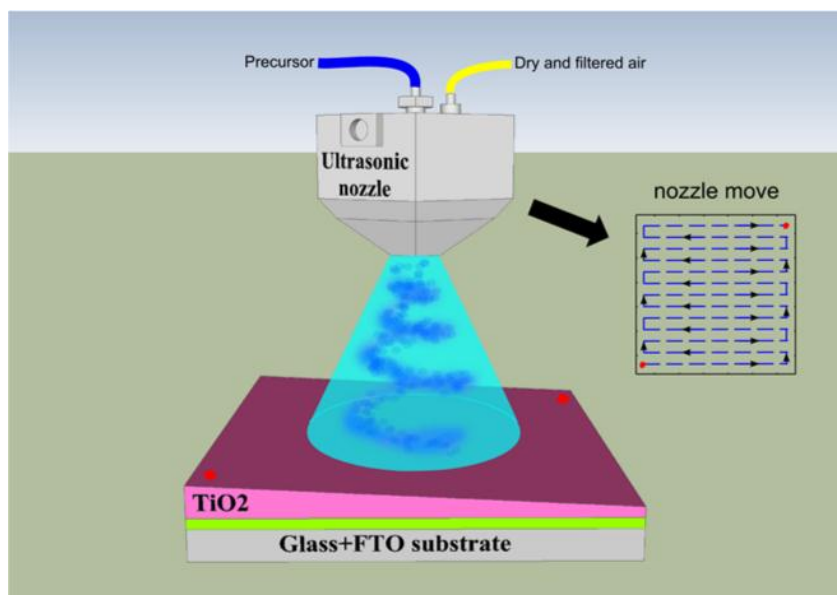

**Figure 5.** Spray pyrolysis deposition process of a linear gradient $TiO_2$ thin film.

## 3.3 Design and Fabrication of $TiO_2$ / ($CuO$-$NiO$-$In_2O_3$) hetero junction solar cells with different back contacts

$TiO_2$ / ($CuO$-$NiO$-$In_2O_3$) cells are a new type of hetero junction solar cells that were not investigated yet. Therefore we've investigated, in detail, combinatorial $TiO_2$ / ($CuO$-$NiO$-$In_2O_3$) hetero junction device libraries for a comprehensive study of the composition of $CuO$-$NiO$-$In_2O_3$ absorbers and their effect on solar cell performance using Ag and Au as metal back contact materials. As a reference for $TiO_2$ / ($CuO$-$NiO$-$In_2O_3$) solar cell efficiency studies, we compared their performance to a binary $TiO_2$ / ($Cu$-$Ni$-$O$) solar cell library.

The FTO substrate served as a transparent conducting front electrode for $I$–$V$ characterization. In order to provide a good electrical contact for the measurement system, the $TiO_2$ / $Cu$-$Ni$-$O$ and $TiO_2$ / ($CuO$-$NiO$-$In_2O_3$) layers, which are close to the library edges, were mechanically removed using a diamond pen, followed by ultrasonic soldering (MBR Electronics) of a thin frame of soldering alloy around the device library. The back contacts, which are grids of 13 x 13 metal patches, were deposited by sputtering or by thermal evaporation onto the $Cu$-$Ni$-$O$ and $CuO$-$NiO$-$In_2O_3$ layers using a shadow mask. Each contact patch had a diameter of 1.8 mm which corresponds to an area of ~0.026 $cm^2$, and defined the solar cell area. The rotation of the shadow mask by 90°, 180° and 270° allowed for the deposition of four back contact materials on a grid of 13 x 13 contact points (see Figure 6). The formerly mentioned contact materials, Au and Ag, were deposited as patches on the $CuO$-$NiO$-$In_2O_3$ absorber layer and photovoltaic activity of the cells was measured. An additional two contact patches (also made of Au and Ag) were deposited onto the NiO hole conducting layer grown on the $CuO$-$NiO$-$In_2O_3$ absorber. The homogeneous thickness of the NiO thin film is ~70nm.

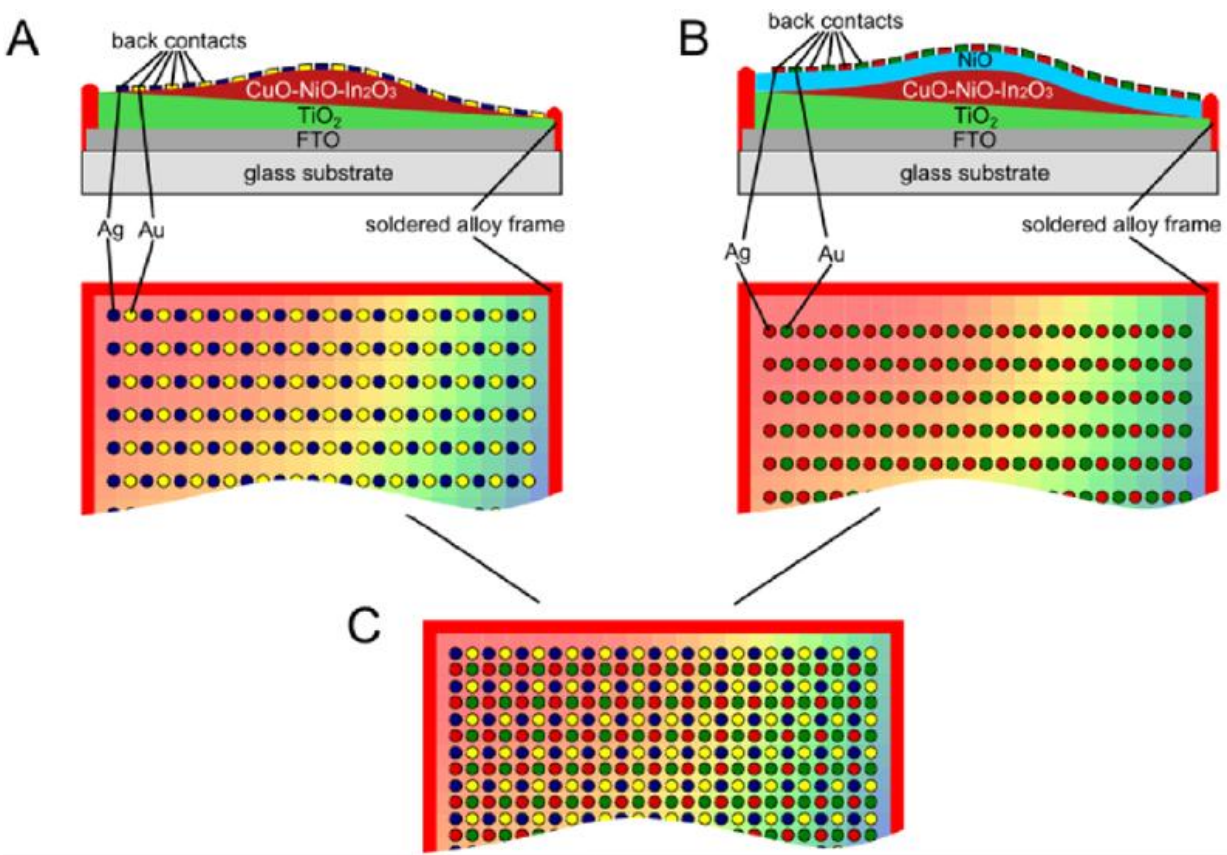

**Figure 6.** Schematic representation of a combinatorial PV device library design, where cross-section and top view of a combinatorial device library are presented: (a) combinatorial device library containing two Ag and Au metal contacts grown onto CuO-NiO-In$_2$O$_3$ thin film; (b) combinatorial device library containing two Ag and Au metal contacts grown onto NiO hole conducting layer; The FTO layer serves as a joint front contact which is electrically connected via an ultrasonically soldered tin alloy frame (red). (c) merged representation of (a) and (b) devices. Four arrays of 13 x 13 round back contacts have been used to investigate the effect of different back contact materials and the NiO hole conducting layer. Each contact patch defines a single PV cell.

# Chapter IV: Characterization methods

This section firstly summarizes the techniques that have been found to be useful for the high throughput characterization of metal oxide thin films and secondly places an emphasis on methodologies that have been used in high throughput material properties evaluations.

## 4.1 X-Ray Diffraction analysis

The performance of thin film metal and metal oxide coatings, in most cases, is determined by the metal/metal oxides crystal structure.

X-ray diffraction is a physical phenomenon as well as an experimental method for the characterization of materials. Polycrystalline samples can take different forms. They can be single-phased or made up of the assembling of crystals of different crystalline phases. The orientation of these crystals can be random or highly textured, and can even be unique, in the case, for example, of epitactic layers[20,21]. The crystals can be almost perfect or, on the contrary, can contain a large number of defects. X-ray diffraction analysis on polycrystalline CuO, NiO, Cu-Ni-O and (CuO-NiO-In$_2$O$_3$) samples enables us to comprehend and even to quantify these characteristics[22]. However, the methods of measuring must be adapted. The quality of the quantitative results obtained greatly depends on the care taken when performing measurements and, in particular, on the right choice of equipment and of the data processing methods used.

The X-ray diffraction technique is suitable for thin film analysis for two main reasons: **1)** the wavelengths of X-rays are of the order of atomic distances in condensed matter, which especially qualifies their use as structural probes. **2)** X-ray scattering techniques are nondestructive and leave the investigated sample or, more importantly, the produced device intact (see Figure 7.).

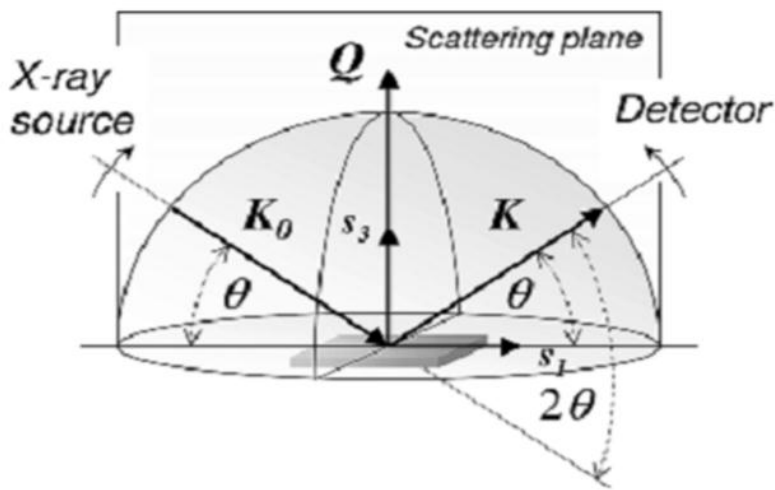

**Figure 7.** Schematic representation of a $\vartheta/2\vartheta$ scan, where $S_3$ - is a substrate normal and $Q$ – is a scattering vector.

The sample positioned in the center of the instrument and the probing X-ray beam is directed to the sample surface at angle $\theta$. At the same angle the detector monitors the scattered radiation. During the scan the angle of the incoming and exiting beam are continuously varied, but they remain equal throughout the whole scan: $\theta_{in} = \theta_{out}$. In X-ray diffraction the angles of incoming and exiting beam are always specified with respect to the surface plane. The quantity measured throughout the scan is the intensity scattered into the detector. The results are typically presented as a function of I($2\theta$) type.

The crystallographic properties of different metal oxides' thin films were evaluated by an X-ray diffraction system called the Rigaku smart lab work station. The system enables the scanning of samples with a relatively large sample size of 72 X 72 mm$^2$. For this purpose we developed certain scanning strategies, which allowed for better thin films characterization (see Figure 8).

Interestingly, for binary and ternary material compositions like Cu-Ni-O or CuO-NiO-In$_2$O$_3$ deposited on glass, the best scanning strategy is matrix points of 4x4 or 7x7. The 4x4 matrix points shown in Figure 8a is the minimal amount of points required to cover the full sample area and it is correlated to the 13x13 matrix points implemented in high throughput scanning systems. For structural analysis of CuO-NiO-In$_2$O$_3$ thin film we used $\Theta$-$2\Theta$ and grazing angle $(2\Theta)$ measurements from 20° to 90°.

For single phase inhomogeneous thin films prepared by pulsed laser deposition (PLD) such as nickel and copper oxides, the conventional and less time consuming way to characterize the material is the use of 7 different points along the maximal thin film thickness changes shown in Figure 8b. The $\Theta/2\Theta$ scan range was performed from 30° to 70° for CuO thin films and from 30° to 80° for NiO thin film.

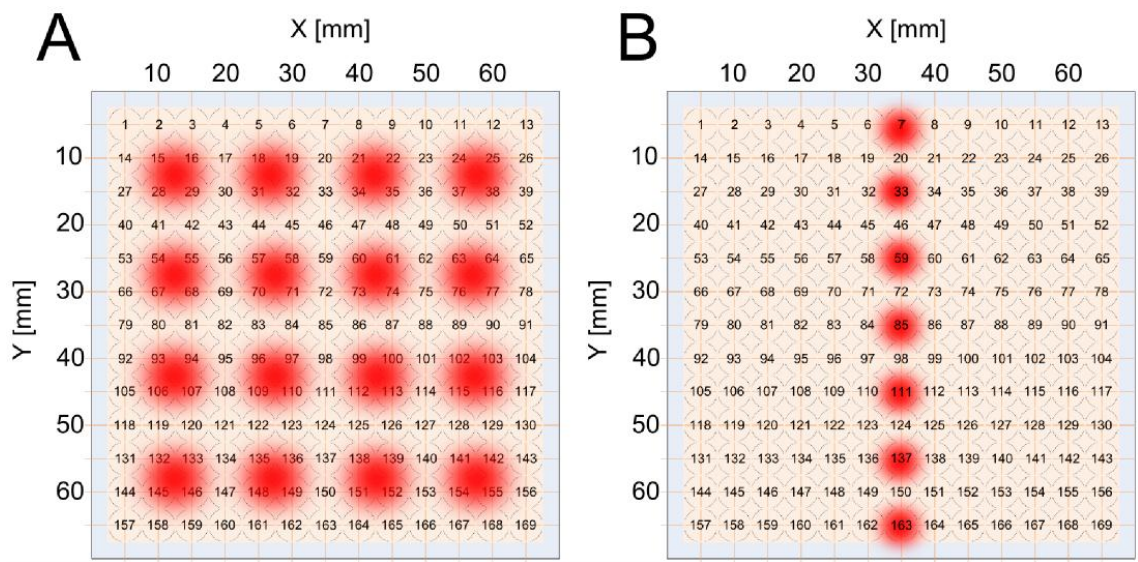

**Figure 8.** Process of scanning thin films for: a) binary, ternary material combinations, b) single material through the maximal thin film thickness changes.

## 4.2 Scanning Electron Microscopy (SEM)

For the morphological metal oxides coatings' structure detection, the scanning electron microscopy method was found to be the most effective technique. Electron microscopy is an extremely versatile technique capable of providing structural information over a wide range of magnification. Scanning electron microscopy (SEM) complements optical microscopy for studying the texture, topography and surface features of powders or solid pieces. In the case of nickel oxide thin films deposited on glass substrates the coating with a thin layer of metal is necessary due to poor electrical conduction, in order to prevent the build-up of charges on the surface of the samples. In addition, the SEM analysis can provide an elemental analysis of sample composition. The system model used in current research is a Helios 600 (FEI).

## 4.3 Energy Dispersive X-ray diffraction (EDX) analysis

The main mode of SEM operation described earlier makes use of the fact that when a sample is placed in the microscope and bombarded with high energy electrons, many things can happen, including the generation of X-rays. These X-rays are characteristic emission spectra of the elements present in the sample. By scanning either the wavelength or the energy of the emitted X-rays it is possible to identify the elements present.

In our research the EDX scanning system was used for the single NiO, CuO and ternary CuO-NiO-$In_2O_3$ material's elemental analysis of 169 sample points. The EDX spectra were obtained by an 80 mm$^2$ X-max detector (Oxford Instruments), which was mounted on a field emission, FEI, Magellan 400L high resolution scanning electron microscope (HRSEM). Moreover the use of 169 EDX spectra points in conjunction with commercially available software (Aztec) allowed us to calculate the thickness of inhomogeneous thin films.

## 4.4 Focused Ion Beam (FIB)

The Focused Ion Beam[23] (FIB) instrument is similar to a scanning electron microscope (SEM), except that the beam that is raster over the sample is an ion beam rather than an electron beam. In FIB instruments, secondary electrons are generated by the interaction of the ion beam with the sample surface and can be used to obtain high-spatial-resolution images. A consistent tenet of any focused beam is that the smaller the effective source size, the more current can be focused to a point. The ion beams are defined by the use of a field ionization source with a small effective source size on the order of 5 nm, therefore enabling the beam to be tightly focused. The size and shape of the beam intensity profile on the sample determines the basic imaging resolution and micromachining precision. Generally, the smaller the beam diameter, the better the achievable resolution and milling precision. The result is a system that can image, analyze, sputter, and deposit material all with very high spatial resolution and controlled through one software program.

The described system was used in our research to create precise cross-sections of samples such as: NiO, CuO, and CuO-NiO-$In_2O_3$ for subsequent imaging and thickness evaluation and correction (see Results and Discussion).

## 4.5 High throughput characterization tools in combinatorial materials science

Combinatorial material science is a time-saving, simple, and convenient method to improve the overall efficiency of testing and screening for new and known semiconductor materials in a sample. This method ensures consistent sample conditions and eliminates the chance for human error. Furthermore, it increases the probability of discovering new materials with the statistical advantage of huge numbers of cells. For this purpose we developed and implemented high throughput tools with a designed screening platform of 13x13 points for screening electrical, optical and photovoltaic properties of metal oxide thin films and all-oxide photovoltaic devices. The high throughput systems used in this work are: optical, conductivity, and IV-point probe scanners.

## 4.6 Optical scanner

The detection of optical properties of metal oxide coatings was performed by using an optical scanner. The scanner is capable of measuring Total Transmission (TT), Total Reflection (TR), and Specular Reflection (SR) with millimeter spatial resolution. The measurements allow the calculation of absorptance, diffuse reflection, light harvest efficiency ($\eta_{LH}$), integrated internal quantum efficiency (in conjunction with the *JV* measurements), layer thickness, band gap and the nature of the band gap (i.e direct/indirect), absorption coefficient, refractive indices, and carrier concentration[24] (see Figure 9). The system was operated at wavelengths of 320 – 1000nm.[25]

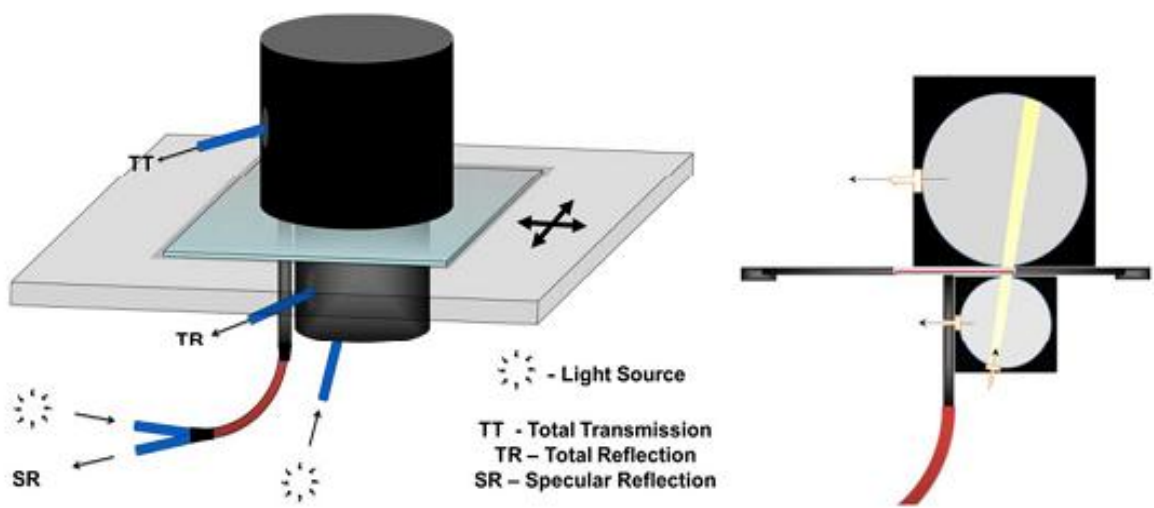

**Figure 9.** Illustration of Optical scanner that measures: Total Transmission (TT), Total Reflection (TR), and Specular Reflection (SR). Left – a schematic 3D drawing. Right – cross sectional view emphasizing the light distribution going to and from a sample.

## 4.7 Electrical conductivity scanner

Conductivity is the essential property of metal oxide film performance, and plays a crucial role in the development of new photovoltaic (PV) systems. Such a property is measured by a high throughput electrical scanning system, which facilitates temperature dependent measurements at different atmospheres for highly resistive samples. The designed instrument is capable of determining conductivity and activation energy values for relatively large sample areas, of about 72x72 mm$^2$, with the implementation of geometrical correction factors. The efficiency of our scanning system was tested using two different samples of CuO and commercially available Fluorine doped tin oxide coated glass substrates. Our high throughput tool was able to identify the electrical properties of both resistive metal oxide thin film samples with high precision and accuracy. The scanning system enables us to gain insight into transport mechanisms with novel compositions such as: Cu-Ni-O and CuO-NiO-In$_2$O$_3$ , and to use those insights to make smart choices when choosing materials for our multilayer thin film all oxide photovoltaic cells. The activation energies of CuO, NiO, Cu-Ni-O and CuO-NiO-In$_2$O$_3$ thin films were measured at a temperature range varying from 27 ℃ up to 300 ℃. In addition, the conductivity measurements with and without light were performed on Cu-Ni-O and CuO-NiO-In$_2$O$_3$ thin film coated glass substrates for evaluation of their opto-electronic performance.

A system was engineered using a Z-arm (Olympus) to lift a set of four probes up and down, an x-y scanning table with 100 mm x-range and 100 mm y-range (Märzhäuser Wetzlar GmbH & Co. KG) onto which a heating stage was mounted to accommodate samples of up to 72 x 72 mm$^2$ and connected to a temperature controller (Eurotherm, model 3216). Then, a home-built linear four point probe head with a constant inter probe spacing of 2.5 mm, achieved using gold plated spring loaded probes (Ingun Prüfmittelbau GmbH), Keithley's voltmeter and current supply, a hermetic Perspex box that is 70 cm in width, length and height for humidity control, and an air temperature and humidity data logger (MRC, model 8808) were incorporated into the system. The system was controlled using a program that was designed in our lab using Labview software. The scanner was tested on FTO and CuO coated glass substrates[26].

## 4.8 IV-point probe scanner

The IV-point probe system was used to measure photovoltaic activity of combinatorial cells in the Cu-Ni-O and CuO-NiO-In$_2$O$_3$ libraries with 4x13x13 points, for a total of 676 unique cells. The current-voltage scans were performed in dark and under light illumination of AM1.5G (one sun). The scanner enables automatic analysis, which classifies *JV* curves by their nature, to photovoltaic or ohmic behavior. Each combinatorial cell was described by the following 6 parameters: 1) 1 sun short circuit photocurrent ($J_{sc}$), 2) open circuit voltage ($V_{oc}$), 3) maximum power point ($P_{max}$), 4) fill factor (*FF*), 5) shunt resistance ($R_{sh}$), and 6) series resistance ($R_s$)[25].

# Chapter V: Results and discussion

The novelty of my research encompasses the: 1. Design and construction of a high throughput electrical scanning system, 2. Characterization of different metal oxides using the designed system, 3. Correlation of the metal oxide properties characterized by the electrical scanning system to the photovoltaic performances of solar cells based on those materials.

## 5.1 Design and construction of a high throughput electrical conductivity scanning system

Characterization of metal oxides' electrical properties provides a better understanding of conduction mechanisms, which results in the efficient use of semiconductors in solar cells. This is why we chose to focus our research on the construction of a new high throughput scanning system, which enables rapid screening of materials' properties and high precision measurements in metal oxide sample characterization. The construction of a high throughput system started with selecting four point probes, which match system requirements such as appropriate scanning area, and high sensitivity, to enable analysis of highly resistive samples. We equipped the system with two outer probes connected to a current supplier (Keithley Model 2400) and two inner probes connected to a voltmeter (Keithley 6517A Electrometer/High resistance meter, Figure 10) which enabled us to detect high thin film resistances up to ~7GΩ.

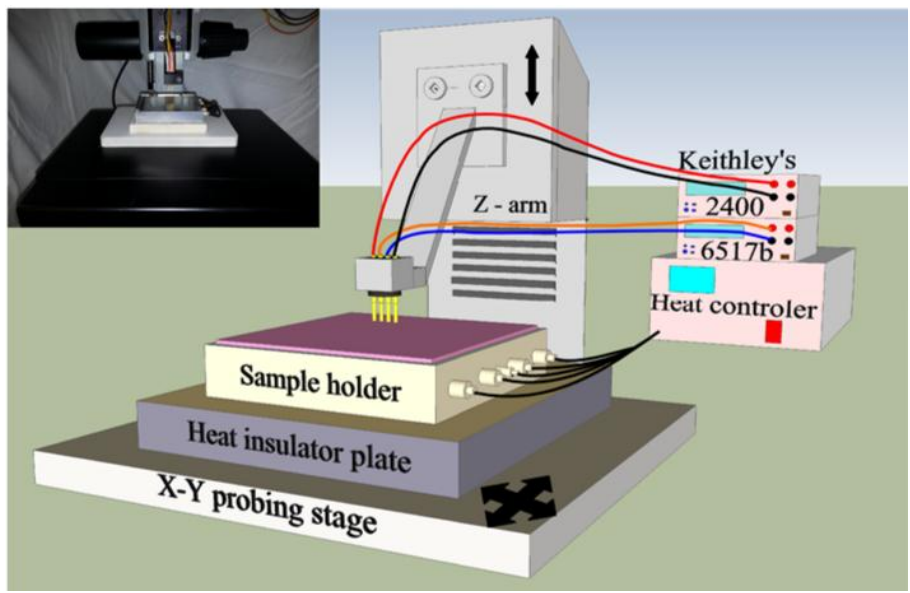

**FIG 10.** Illustration of scanner system which consists of: four point probe head, sample holder, four equally dispersed heaters with a thermocouple in the middle, heat insulator plate, X-Y probing stage, and Z – arm. A picture of the high throughput electrical scanner is shown on the top left corner of the image.

The heating stage was designed to enable homogeneous heat distribution through the sample interface by using four equally dispersed heaters in parallel, inside the sample holder and a thermocouple (see Fig 10). The homogeneity of the heat distribution among the sample holder was detected using an IR camera at two sample holder temperature values of 50°C and 200°C, which were measured by a thermocouple located inside the sample holder plate. The temperature readings on the sample holder area shows a deviation of 0.6 °C throughout the holder, when the sample holder was heated to 200 °C.

A rectangular array of 13 rows and 13 columns, accounting for a total of 169 semiconductor thin film points, was employed for two-dimensional sample mapping for resistance evaluation, which is shown in Figure 11(a).

The resistance of thin films was evaluated using the four point probe technique, implemented in the scanner. This method practically eliminates measurement errors due to contact resistances between the probes and the sample, which usually exist in the two point probe measurement technique. Moreover this technique is widely used in the semiconductor industry to monitor the production process[27, 28] and provide information on the various process steps[29, 30] and active carrier concentration of doped semiconductor films[31].The resistance measurement results were then used for conductivity value calculations. Conductivity, as opposed to resistance, is an intrinsic physical property, independent of the particular shape or size of the sample and important for the interpretation of solar cell performance. The resistivity and conductivity of FTO and CuO thin films were calculated using equations (1) and (2).

$$\rho = \frac{\pi t}{\ln 2} * (V/I) * F \quad , s >> t \qquad (1)$$

$$\sigma = 1/\rho \qquad (2)$$

Where $\sigma$ - is conductivity and $\rho$ - is resistivity which is a function of the measured film thickness - t, current flowing through the sample – I, voltage detected across the sample - V, geometrical correction factor - F and the integration constant - $\pi/\ln(2)$ which is obtained when probes are uniformly spaced and the space is larger than film thickness.

Geometrical correction factors were applied for conductivity values observed for the sample[32]. The geometrical correction factor expressed in equations (3), (4), and (5) is based on a theoretical model for rectangular samples. The model was found to effectively correct the conductivity/resistivity edge values (see Figure 11). The geometrical correction factor $F$ is evaluated by the superposition of two geometrical correction factors: 1) Correction factor $F_\perp$ applied to a sample when the scanner's probes are positioned perpendicular to the Y axis, 2) Correction factor $F_{||}$ applied to a sample when the scanner's probes are positioned parallel to the X axis.

$$F = F_\perp * F_{||} \qquad (3)$$

$$F_\perp = \cfrac{1}{1 + \frac{1}{2ln2}ln\frac{\left(\frac{L_2}{S}+2\right)\left(\frac{L_2}{S}+1\right)}{\left(\frac{L_2}{S}+\frac{5}{2}\right)\left(\frac{L_2}{S}+\frac{1}{2}\right)}} \qquad (4)$$

$$F_{||} = \cfrac{1}{1 + \frac{1}{2ln2}ln\frac{(L_1/S)^2+1}{(L_1/S)^2+1/4}} \qquad (5)$$

$L_1$ and $L_2$ are the distances from one of the outer probes to the sample's X and Y edges respectively, S represents the constant distance between probes (see Figure 11(b)).

However, the accuracy of the measured resistance values (obtained from the conductivity scanner) is affected by the correlation between the thickness of the thin film's coating and the spacing between the scanner probes. In the cases where this correlation was above 40%, the conductivity calculations are indicative of undesired bulk materials and additional correction is needed. In our system, due to the relatively wide 2.5 mm distance between the probes, the additional correction factor will always be less than 40% and result in a factor value of 1. Figure 11(c) shows the map of geometrical correction factor values for rectangular samples which decrease along the X − axis ,due to the inequality in edge distances where $L_1 < L_2$ (see Figure 11(b)). The high correction factor values obtained from the central part of the sample are found to be 0.99 − 1, while the lowest values were detected at the edge of the sample and vary from 0.89 to 0.97.

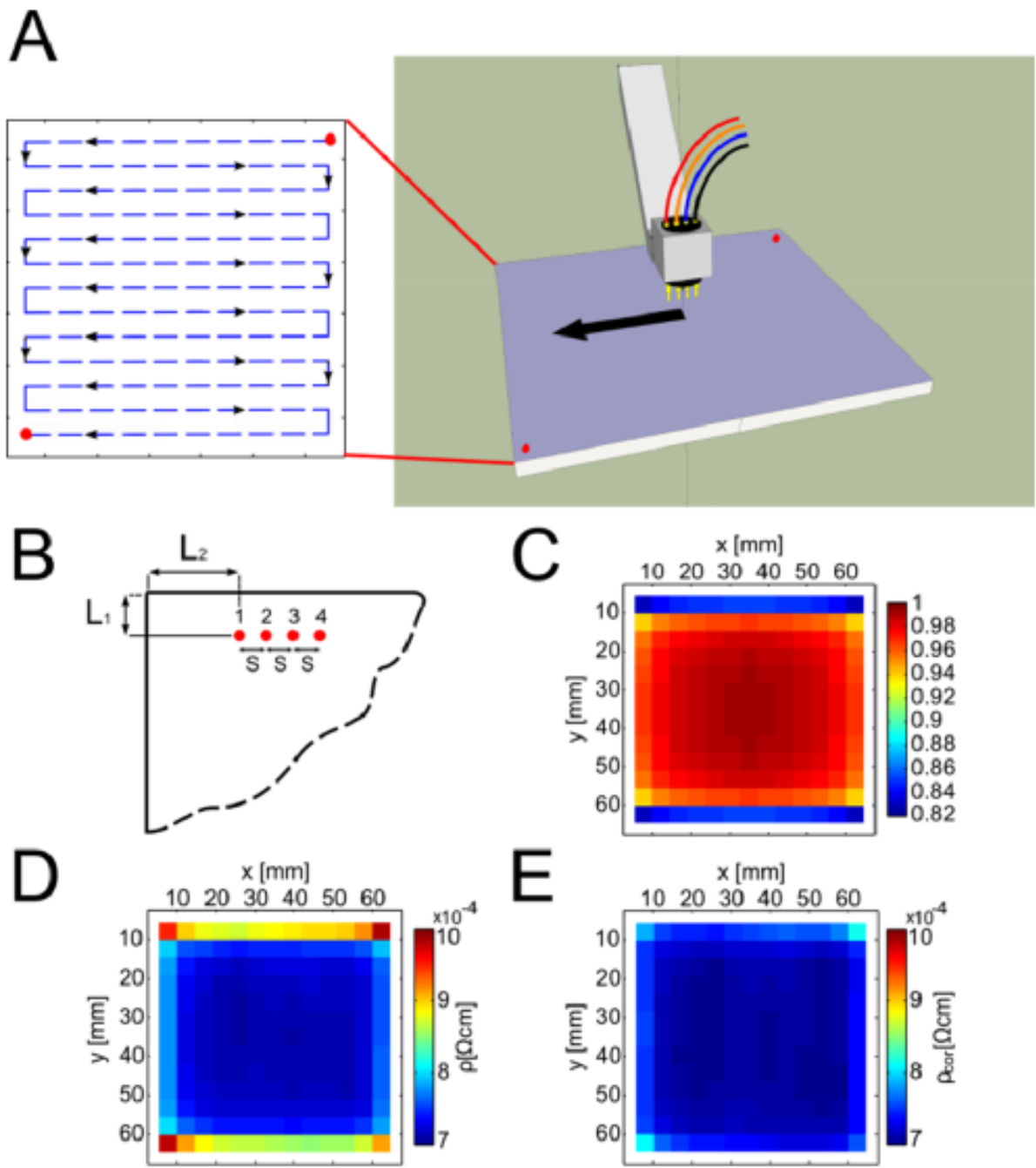

**FIG 11**. (a). Process of sample scanning, where the probes are lifted up, the X-Y scanning table moves to the next position and the probes come down to the sample surface again, (b). An illustration of probes' position at the edges of the sample, (c). Correction factor map calculated using equation (3), (d) FTO resistivity as measured at 47°C, (e). Corrected resistivity of FTO sample at 47°C; a product of (c) and (d).

To demonstrate the performance of our system on metal oxide thin film coatings we chose to use a copper oxide (CuO) semiconductor. First, the CuO film thickness was measured.

Conductivity is calculated using resistance and thickness values, therefore accuracy in thickness measurements is of paramount importance. Figure 12(a) shows the map of thickness gradient values for high quality copper oxide coating measured by an Energy Dispersive X-ray (EDX) scanning system and confirmed by a cross section image taken using a Focused Ion Beam (FIB). The conductivity of CuO, like other semiconductors, increases with temperature due to the extension of its Fermi function which increases the number of charge carriers. Understanding temperature dependent conductivity is critical for understanding device performance, as well as for extracting fundamental information on conduction mechanisms and activation energy for carrier transport. The activation energy was evaluated using equations (6) and (7) in which the slope of a semi log plot of $\ln(\sigma)$ versus $T^{-1}$ equals -$E_a/k$ (Arrhenius behavior)

$$\sigma = \sigma_0 \exp(- Ea/kT) \qquad (6)$$

$$\ln(\sigma) = \ln(\sigma_0) - \frac{E_a}{kT} \qquad (7)$$

Where $\sigma$ - is the conductivity of the film, $E_a$ – is the activation energy, and $\sigma_0$ – is the pre exponential constant.

Figure 12(b) shows the map of corrected conductivity values which vary from $1 \times 10^{-4}$ to $4.5 \times 10^{-4}$ [S/cm]. Such small variations in conductivity are consistent with the XRD data (see Figure 12(c)), predominantly showing the formation of CuO, which appears in *cubic* structures, for the entire sample and an increase in planes orientation with increase in film thickness. The conductivity values higher than $4.5 \times 10^{-4}$ [S/cm] are considered a measurement error due to the deformation of the film at thicknesses lower than ~30nm, where pressure has a major effect and deforms the thin film[28].

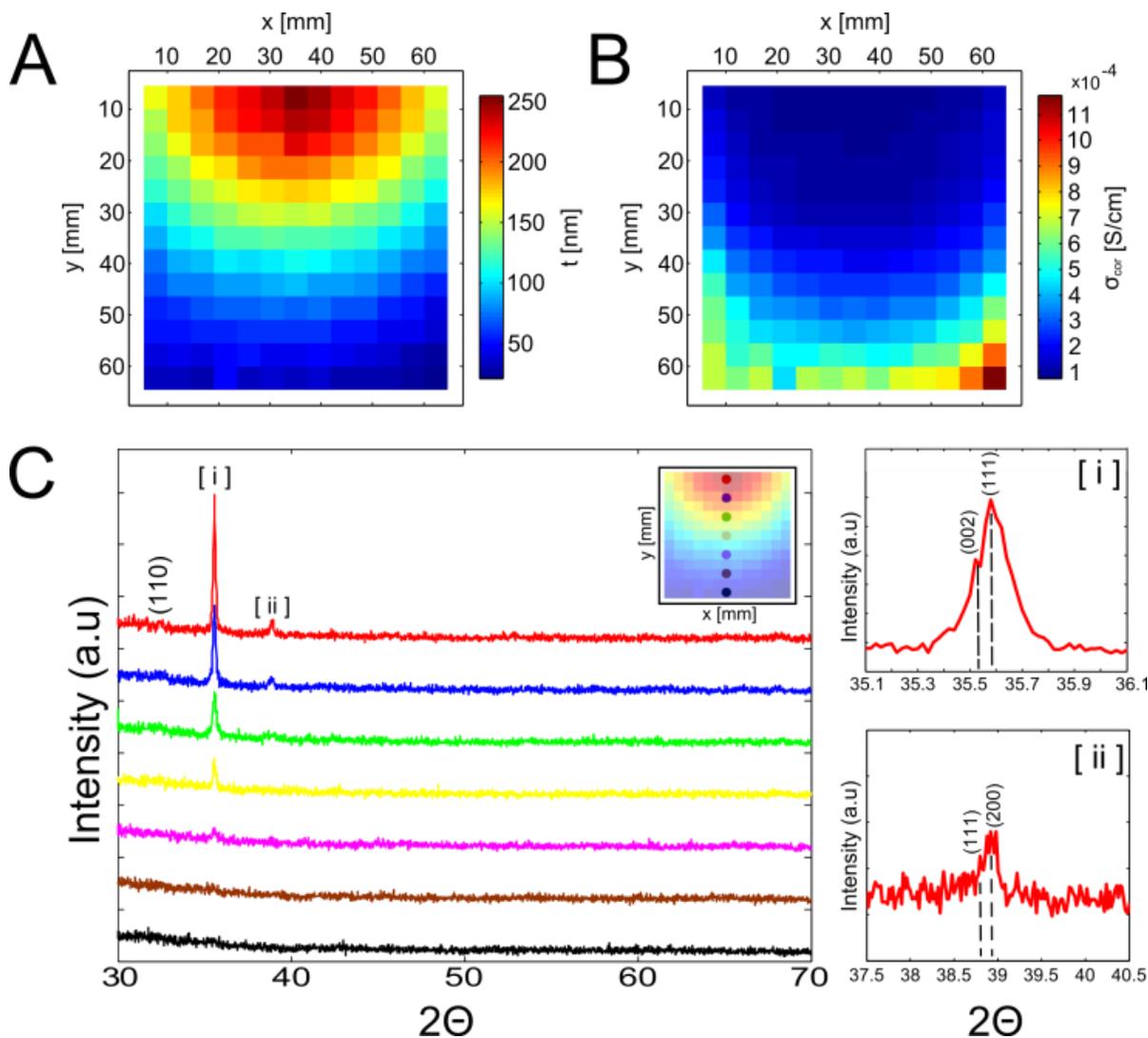

**FIG 12**. (a) A 169 point, 2D plot of the CuO thickness gradient in the measured film. (b). Conductivity of the CuO film at 47 ºC, including geometrical factor correction. (c). $\vartheta/2\vartheta$ XRD measurements of the CuO thin film. The graphs' colors correspond to the measured points indicated in the upper right insert.

Figure 13(a) shows Arrhenius plots for temperatures varying between 28°C and 247°C revealing two regimes of activation energies that can correspond to two main conduction mechanisms: 1. Electron hopping from one trap to the other at a low temperature range and 2. Intrinsic conduction is dominant at high temperature zones[33]. Therefore, the activation energy for the entire sample was calculated and is presented for the two regions (Figures 13(b) and 13(c)). For temperatures between 28°C and 167°C the activation energy remains identical throughout the sample area with a value of 0.28eV and an error of $k_bT \cong 0.026$eV. In the temperature range of 187°C to 247°C the activation energy is not uniform, increasing to values between 0.45eV and 0.65eV.

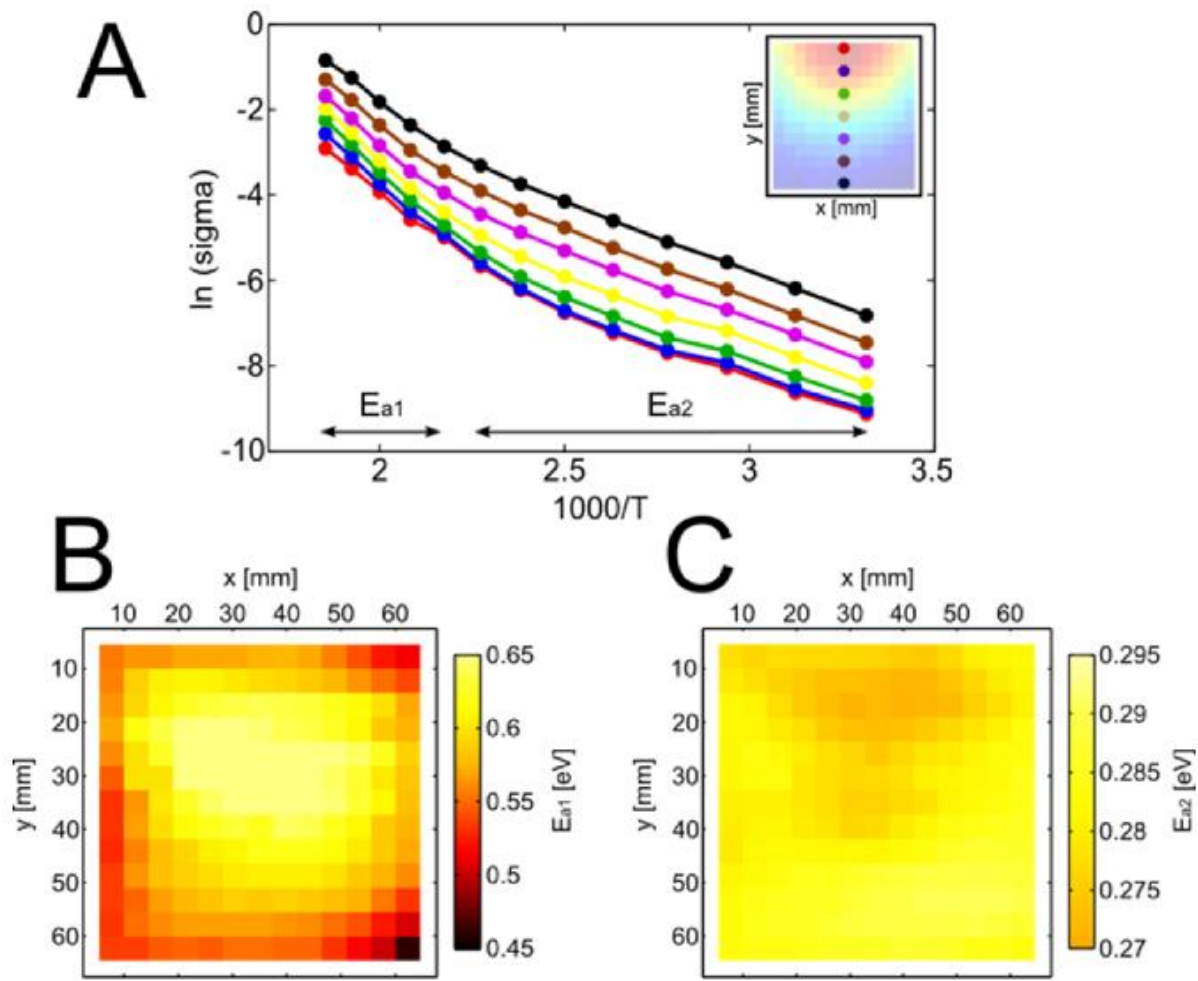

**FIG 13**. (a) Arrhenius plots along gradient. (b), (c) Maps of activation energy calculated at temperatures range of 167°C-247°C and 28°C- 167°C respectively.

## CONCLUSIONS

We demonstrated the development of a resistivity, conductivity and activation energy mapping instrument. The effectiveness of the instrument was probed on cuprous oxide thin films with inhomogeneous thicknesses and a commercial FTO homogeneously coated sample.

The mapping instrument was designed to detect high thin film resistances at the range of ~7GΩ. It contains geometrical correction factors thereby providing high accuracy including the sample edge. The system can operate under different atmospheric environments in a temperature range of 27°C to 300°C. The designed instrument provides a means for determining the mechanisms of electrical conduction in the sample mainly in high throughput studies.

## 5.2. Electrical and optical characterization of inhomogeneous NiO coatings

Nickel oxide is one of the most promising, low-cost, polarizable materials[34] that exists since it is characterized as having excellent durability, electrochemical stability, high transmittance of ~80-90% at the spectrum range of 400-900 nm and no bulk redox or self-discharge reactions in the range of applied potential. It is known as a p-type semiconductor[35,36,37] and therefore should be a suitable material for use as a holes transport layer in All-Oxide solar cells. In our current research we investigate the structure, electrical and optical properties of nickel oxide coatings deposited by pulsed laser deposition (PLD) on glass substrates under different oxygen pressures, temperatures, applied laser energies, number of pulses and target substrate distances. A powerful motivator for this line of research is the observation that NiO thin film's structural, electrical and optical properties can be precisely controlled and optimized[5] to specific orientations, conductivities, activation energies, absorptances and band gap values using high throughput tools[25,38,39,26] described earlier in Chapter IV of Characterization methods. Moreover, we demonstrate the general strategy of high throughput sample analysis where the pre-screening process (which is described in detail in the Chapter III of Experimental Section) plays a crucial role for the identification of our main PLD deposition parameters. We then show a detailed analysis of the structural, electrical and optical properties of an optimized NiO coated sample with a large surface area.

**Thickness analysis**

The film's thickness is an important parameter used in high throughput analyses. The thickness analysis of NiO coated samples was performed using an FIB system and presented in Figures 14(a) - (g), which show the cross-section images of seven points across the film gradient. The point located at the NiO sample center (see Figure 14d) serves as an example point of band gap evaluation in optical analysis. Accurately evaluating the sample's thickness is necessary in order to achieve high accuracy in resistivity, conductivity, and band gap values of the film's calculations. The determined thickness values of the sample used in X-ray diffraction quantitative analysis are 200; 185; 145; 90; 70; 40; and 20 nm respectively. The linear behavior in the thicknesses of the film was observed using the, EDX and FIB techniques, which were used for thickness data analysis (see Figure 14 (h-i)).

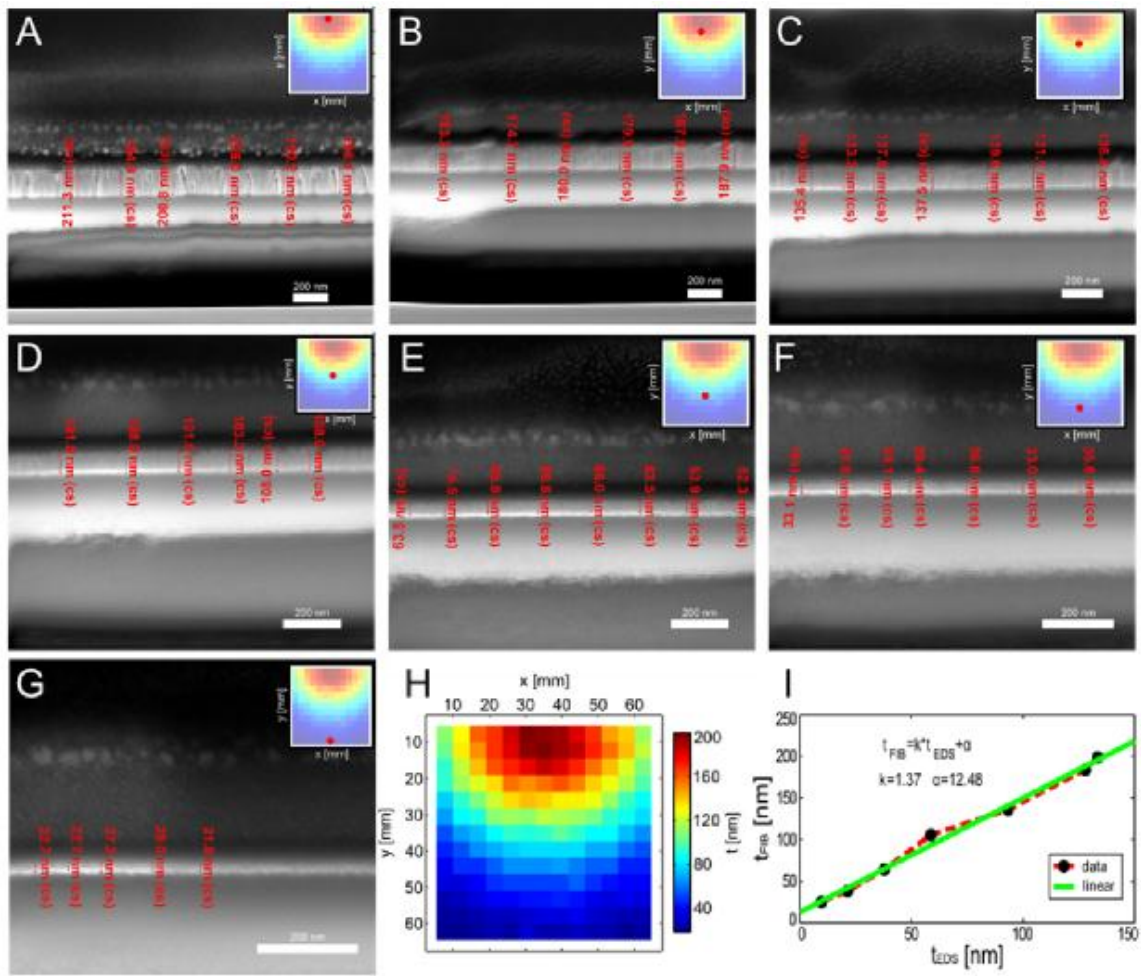

**Fig 14.** a-g) Seven FIB cross section points of NiO coating deposited on glass substrate, h) Thickness profile of 169 data points corrected using FIB and EDX systems, where thickness corrections were made using linear behavior shown in i).

Following this analysis, and using the rules of linear behavior between these two techniques, the thickness was corrected using equation (8):

$$t = t_{EDX} * k + \alpha \qquad (8)$$

Where $t$ - is the final thickness of the nickel oxide film, $t_{EDX}$ – is the thickness measured by the Energy Dispersive X-ray (EDX) scanning system, $k$ – is a linear factor, and $\alpha$ – is the line offset. Figure 12(d) illustrates the thickness values' distribution of high quality gradient nickel oxide coating.

Two SEM images of Fluorine doped Tin oxide substrate, and NiO thin film deposited on the Fluorine doped Tin oxide substrate are presented in Figure 15 (a-b) to show the independence of substrate structure.

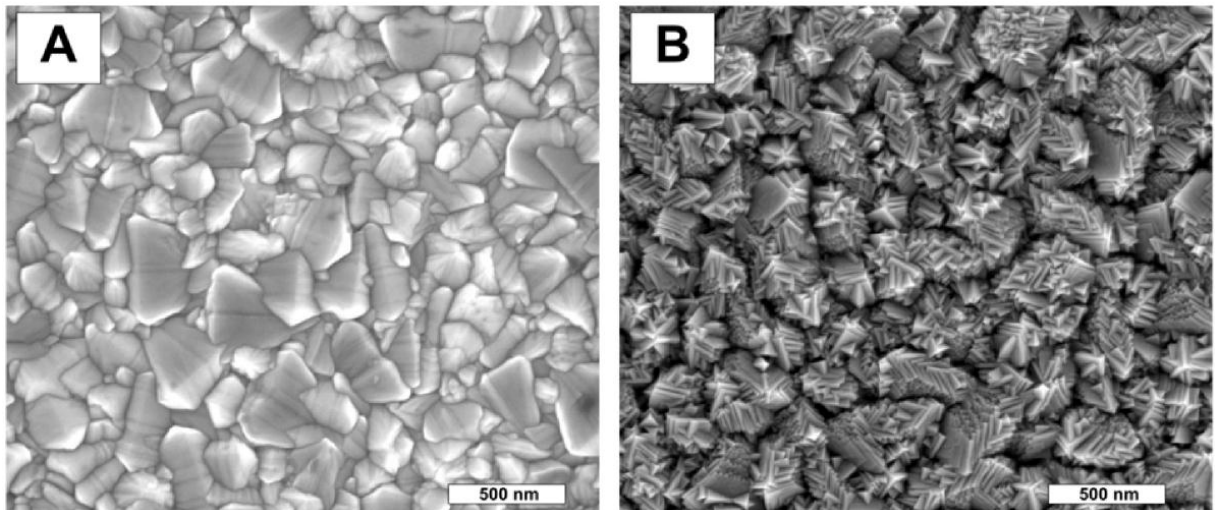

**Fig 15.** a) SEM image of Fluorine doped Tin oxide thin film coated onto a 2.2 mm thick glass substrate, b) SEM image of NiO deposited on Fluorine doped Tin oxide thin film.

**Analysis of electrical properties**

For electrical analysis of the NiO thin film we used an in-house designed high throughput electrical resistance scanning system (see reference 22 and section 5.1). The electrical resistivity was calculated and corrected using a geometrical correction factor map. The geometrical correction factors described earlier (see Figure 11 and 16(b)) and calculated using equations (3), (4), and (5).

As for CuO thin film, the resistivity values of the NiO thin film at the edges of the sample fluctuate due to the geometry of the sample and distance of the probes from sample edges. The conductivity of the NiO thin film was calculated using corrected resistivity values and presented in Figure 16a. We observed only small changes in conductivity values for the entire area of the sample from $0.5*10^{-4}$ [S/cm] to $1.5*10^{-4}$ [S/cm]. The conductivity values above 1.5*10⁻⁴ [S/cm] are considered to be a measurement error due to the deformation of the film at thicknesses lower than ~40nm.

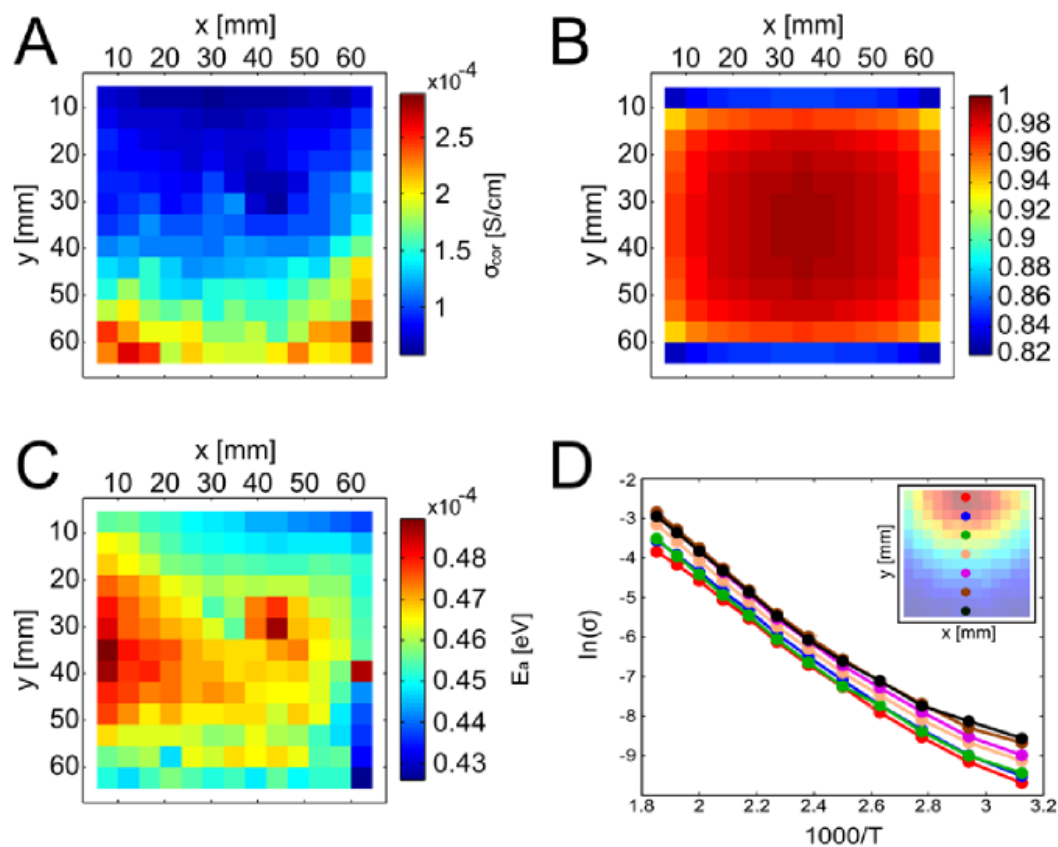

**igure 16.** a). Corrected conductivity of nickel oxide coating measured at a temperature of 47°C and 8.1% humidity, b). Correction factors map, c). Activation energy map of 169 points, d). Arrhenius plot.

The conductivity measurements performed under different temperatures enable the determining of the transport mechanism of large surface area NiO coatings. The Arrhenius plot, which is shown in Figure 16(d) corresponds to small polaron behavior, where the NiO lattice fluctuates due to the movement of electrons. The activation energy map with values varying from 0.43eV to 0.49eV was calculated using the Arrhenius slope (see Figure 16(c-d)). The differences in activation energy, which exceed an allowed error of $k_b T \cong 0.026eV$, correspond to the changes in texture factor according to the results reported in the section titled *Structural analysis via X-Ray Diffraction*.

## Structural analysis via x-ray diffraction

NiO thin film was deposited by PLD on a glass substrate and its crystalline nature was confirmed by X-ray diffraction (XRD) patterns which are presented in Figure 17. All seven XRD patterns show a pure polycrystalline nature with no sign of a detectable second phase. XRD patterns reveal that the major diffractions come from the $2\theta =37.25°$ and $79.4°$, corresponding to (111) and (222) planes, and indicate the presence of oriented crystallites in the (111) direction. Careful examination of the peak corresponding to $2\theta =79.4°$ shows two humps (Figure 17) which could be due to either $K_{\alpha 2}$ radiation from cubic (PDF card No: 01-071-1179) or tetragonal (PDF card No: 00-044-1159) crystal structures.

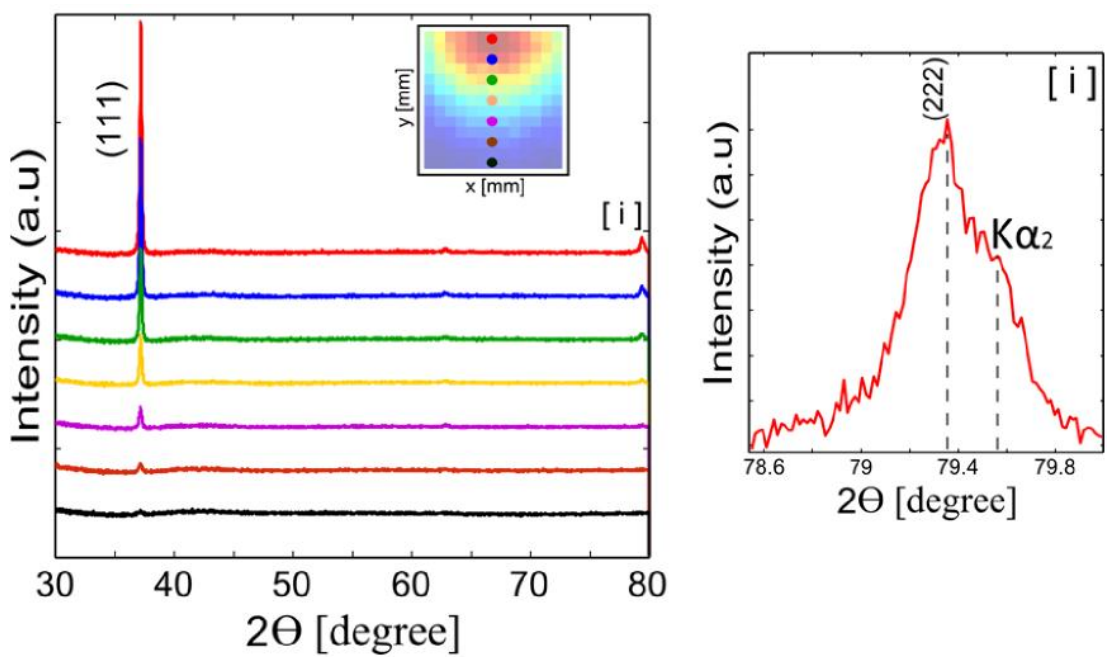

**Fig 17.** Left: $\vartheta/2\vartheta$ measurements of gradient NiO thin film performed without monochromator. The picture at the right presents the magnified image of the X-ray diffraction pattern at 79.4°.

To resolve this uncertainty, we have measured the same points using a monochromator (Ge 220). Figure 18a shows the XRD pattern with monochromator for seven different points along the thickness gradient as shown in the inset. After measuring with a monochromator the second hump disappeared suggesting it's originating from $K_{\alpha2}$ radiation in cubic structure (Fm3m, PDF card No: 01-071-1179) of the NiO film.

From the above XRD data we have calculated the degree of orientation in the (111) direction known as texture factor ($T$) by using the following formula

$$T = \frac{\frac{I^m}{A_{\theta2\theta}}}{I^{ICDD}} \frac{\sum I^{ICDD}}{\frac{\sum I^m}{A_{\theta2\theta}}} \qquad (9.1)$$

$$A_{\theta2\theta} = 1 - exp^{(-\frac{2\mu t}{sin\theta})} \quad (9.2)$$

Where $A_{\theta2\theta}$ – is the absorption factor, which is a function of the diffraction angle $\theta$, $\mu$ is the linear attenuation coefficient, which is dependent on the material's property (calculated for the NiO), and $t$ is the film thickness, measured at specific points. $T$ (texture factor) is dependent on three things: $I^{ICDD}$, intensity data from the ICDD (cubic (fcc) NiO), $I^m$ integral intensity of the measured reflection, and $A_{\theta2\theta}$, the absorption factor. The texture factor, calculated using equations 9.1 and 9.2, indicates a high degree of orientation for five points along the thickness gradient. Two points at the bottom edge of the sample where the thickness of the film is less than ~40 nm (Figure 14 h) are not presented due to the uncertainty of the X-ray diffraction measurements.

Figure 18b show the plot of activation energy ($E_a$) with texture factors ($T$). Activation energy decreases with the increase of the texture factors, revealing that there is a strong dependence of activation energy on texture factor for the pulsed laser deposited NiO thin film.

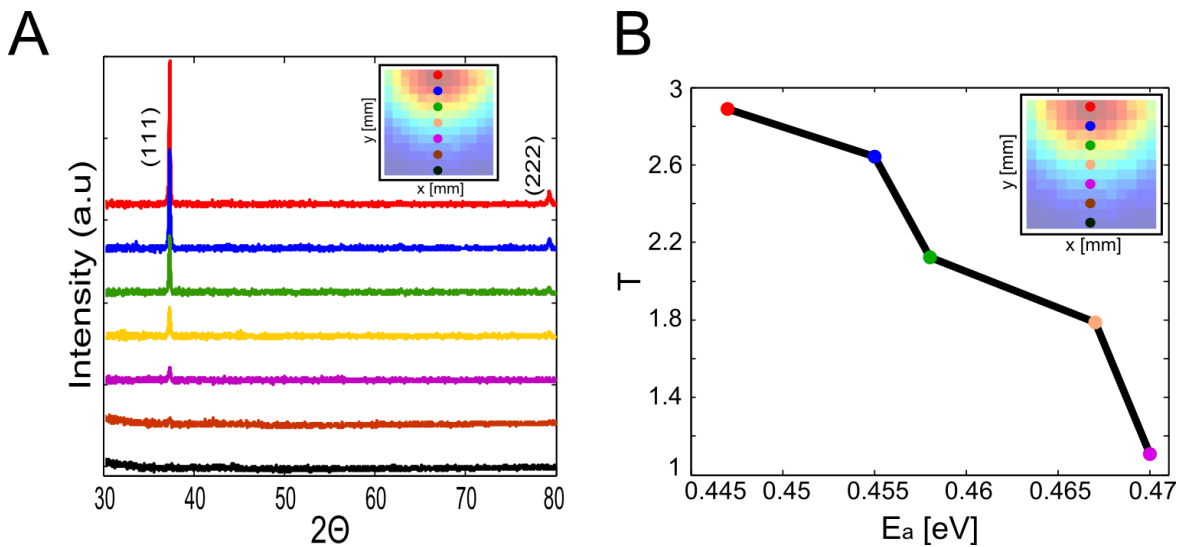

**Figure 18**. a) $\vartheta/2\vartheta$ measurements of gradient NiO thin film with deposition parameters described previously in the PLD section (T=400°C, z=77mm, flow=65sccm, $P_{total}$=1.32E-01Torr). b) Activation energy versus texture factor (see *Electrical analysis*). Two points at the bottom edge of the sample where the thickness of the film is less than ~40 nm are not presented due to the limitation of the X-ray diffraction system to detect peaks of NiO thin films below ~40nm.

## Optical analysis

To analyze the optical band gap values of the 169 scanned sample points we relied on the relation presented by equation (10.1)

$$\alpha h\nu = A(h\nu - E_g)^{1/2} \qquad (10.1)$$

Where $A$ is a constant, $h\nu$ is the photon energy and $\alpha$ is the absorption coefficient.

The absorption coefficient $\alpha$ was derived using equation (10.2) from the absorptance spectra, calculated using equation (10.3) in a wavelength range from 320 -1000 nm.

$$\alpha(x, y, \lambda) = t(x, y)/(-\log(1 - A(x, y, \lambda))) \qquad (10.2)$$

$$A(x, y, \lambda) = 1 - TT(x, y, \lambda) - TR(x, y, \lambda) \qquad (10.3)$$

Where $A(x, y, \lambda)$ is the absorptance, $\alpha(x, y, \lambda)$ is the absorption coefficient, $TT(x, y, \lambda)$ and $TR(x, y, \lambda)$ are the total transmission and the total reflection (respectively) for each point with coordinates x and y and varying wavelengths - $\lambda$ and thin film thickness - $t(x, y)$ at each point with coordinates x and y.

Figure 19d shows the plot of $(\alpha h\nu)^2$ vs photon energy h$\nu$ for a point located at the center of the sample. For 169 points the nature of the plot indicates the existence of direct optical transition. The band gap, $E_g$, has been graphically determined by extrapolating the linear portion of the plot to the energy axis h$\nu$ using the r-squared values map presented in Figure 19c. The band gap values for the inhomogeneous coating are around 3.54eV with a fluctuation of 0.07eV, which is in the range of reported results[41] of $3.15 - 3.8$eV.

Figure 19b shows the absorptance profile at a selected wavelength of 320nm, where the NiO film maximally absorbs light.

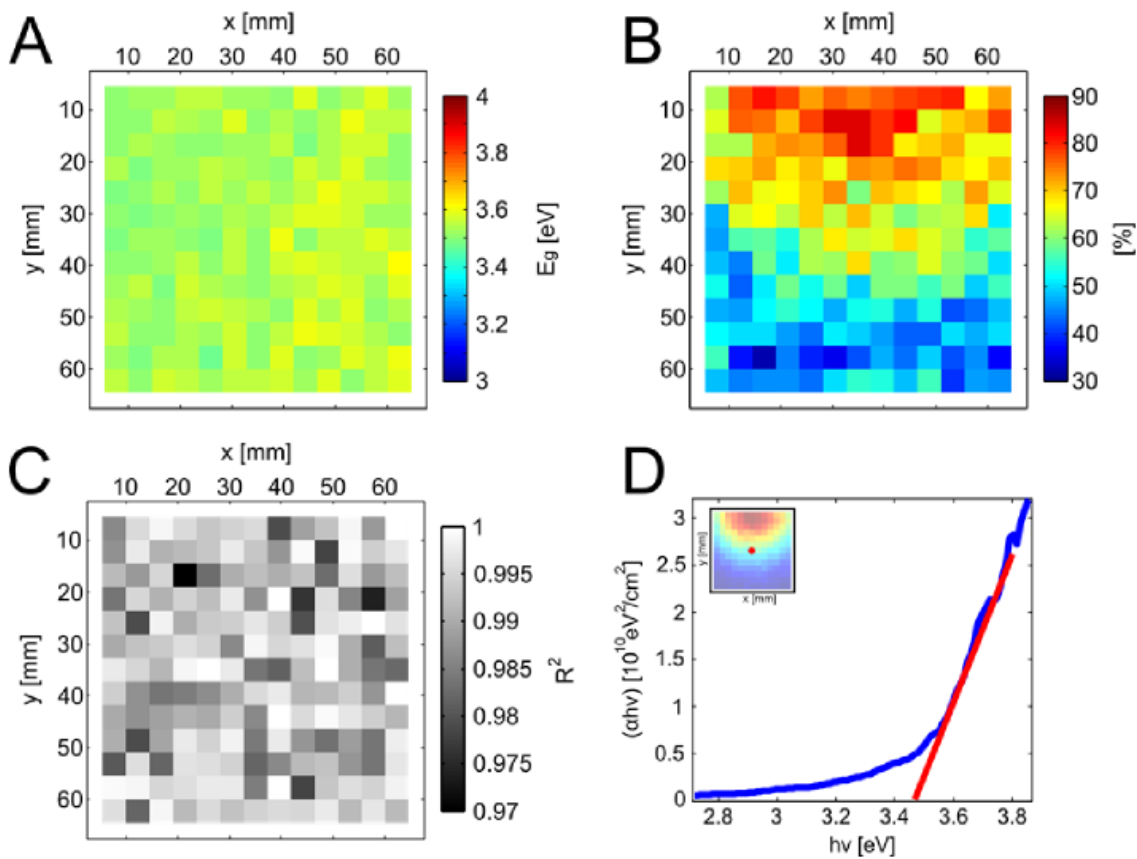

**Figure 19**. a) Band gap map of 169 inhomogeneous NiO thin film points , b) Absorptance map measured at 320 [nm], c) R-squared map used to evaluate the linear extrapolation of the $(\alpha h\nu)^2$ versus photon energy ($h\nu$), d) Plot of $(\alpha h\nu)^2$ versus photon energy ($h\nu$) for a point located at the NiO sample center.

## Conclusions

The high throughput approach for the rapid discovery, study, and optimization of nickel oxide thin films was used in NiO thin film characterization in our current research work. The most effective nickel oxide properties were found during the early stage of the study. The nickel oxide thin film with the most desirable electrical, optical and structural properties became the object for improvement of performance for use in solar cells. Correlation between electrical, optical and structural properties with pulsed laser deposition parameters was observed. Thus, nickel oxide thin films have been found as a promising class of metal oxide material for use as a hole-selective contact for solar cells based entirely on metal oxides.

## 5.3. Combinatorial discovery of CuO-NiO-In$_2$O$_3$ photo absorbers for All-Oxide solar cells

Light absorbers play a critical role in photovoltaic cells, where the incident photons are absorbed and energetic electrons and holes are generated and transported in order to provide current to an external circuit. To evaluate the performance of a light absorber, an asymmetric junction solar cell is typically introduced to measure charge separation and voltage generation properties. However, such measurements do not reflect intrinsic semiconductor performance. For instance, the most common figure of merit in these measurements, the short circuit photo-current $(J_{sc})$, could be governed by poor charge separation across the junction even if the semiconductor has good carrier collection efficiency. Therefore, we describe a fundamental study of a ternary CuO-NiO-In$_2$O$_3$ thin film system fabricated on a glass substrate, which was found to be a promising new light absorber material for use in all oxide solar cells. The inspiration for this work was taken from the previous studies of binary CuO-In$_2$O$_3$ metal oxide systems[42,43,44], which were mostly studied and used in solar energy harvesting applications. The CuO-In$_2$O$_3$ binary system is chemically stable under air, has inherently high carrier concentrations, and can be used at significantly higher temperatures than traditional silicon PV cells. For the CuO-In$_2$O$_3$ system two material phases, Cu$_2$In$_2$O$_5$ and CuInO$_2$, were reported, wherein the latter phase has been proven to be of bipolar dopability, namely, it has the capability of being doped to be both a p -type (with Ca) and n -type (with Sn) semiconductor[45,46].

Copper oxide (CuO), a p-type semiconductor, has been considered for photovoltaic applications due to its high conductivity and relatively small direct band gap (1.4-1.8 eV) compared to $In_2O_3$, an n-type semiconductor, but it has never been adopted for major device applications. The difficulty in controlling copper oxides' electrical properties is one of the reasons limiting its practical use in photovoltaic applications.

Relying on the previous studies of CuO and NiO thin films described in this thesis (see sections 5.1 and 5.2), and taking into consideration previously reported studies on binary $CuO-In_2O_3$ material systems, we propose a general strategy for the formation of a new metal oxide ternary $CuO-NiO-In_2O_3$ system for potential use in all oxide solar cells. We show the structural, electrical, and optical characteristics as well as phase composition of the $CuO-NiO-In_2O_3$ thin film prepared by pulsed laser deposition technique at 400 °C.

**Structural properties**

In the mapping of composition-structure property relationships, one piece of information that is of paramount importance is the phase and crystal structure distribution. To obtain a comprehensive phase analysis across composition spreads of complex materials we used agglomerative hierarchical clustering analysis. This type of analysis has been developed by C.Long and I.Takeuchi and employed as a software program (CombiView)[47]. By using a CombiView program we mapped out specific regions of composition space where the crystal structures are identical (see Figure 20(b-c)). Background subtraction and normalization on the data were performed by using hierarchical clustering analysis. A dendogram was used to adjust the number of groups in the sample (see Figure 20d). The similarity between all pairs of spectra was calculated by sorting the spectra into discrete regions, using the Pearson correlation function and the distribution of points was calculated. In order to reinforce the credibility of using the Pearson correlation function for the $CuO-NiO-In_2O_3$ metal oxide system, data mining calculations were performed to test the fitting correlation function. In the data mining process of $CuO-NiO-In_2O_3$ thin, film 6 parameters were used: 1) thickness of the thin film, 2) composition of the film: Cu%, Ni%, In%, O%, and 3) Band gap $(E_g)$. The product of the data mining is presented as a decision tree (see Figure 21). The decision tree algorithm is a classification algorithm that can be used for data mining and as a predictive model. The decision tree algorithm operates by iteratively splitting a dataset characterized by classes and descriptors into smaller subsets. At each step, all descriptors are considered in a search for one that, upon splitting a parent node, would produce the most uniform child nodes. This procedure was repeated until no more splits are warranted either since all compounds within all (terminal) nodes have identical classes or since the gain in uniformity upon additional splits were not statistically significant. In the present study we used the J4.8, a C4 variant algorithm implemented in the WEKA version 3.7.9 software.

A map of XRD spectra as a function of composition is presented in Figure 20b. We observed that no clear transition in the diffraction spectrum (see Figure 20a) was detected for the transfer from the first (blue) group to the last (red) group. This result was interpreted as a presence of multiple crystal structures in the sample, which change slowly with change of composition.

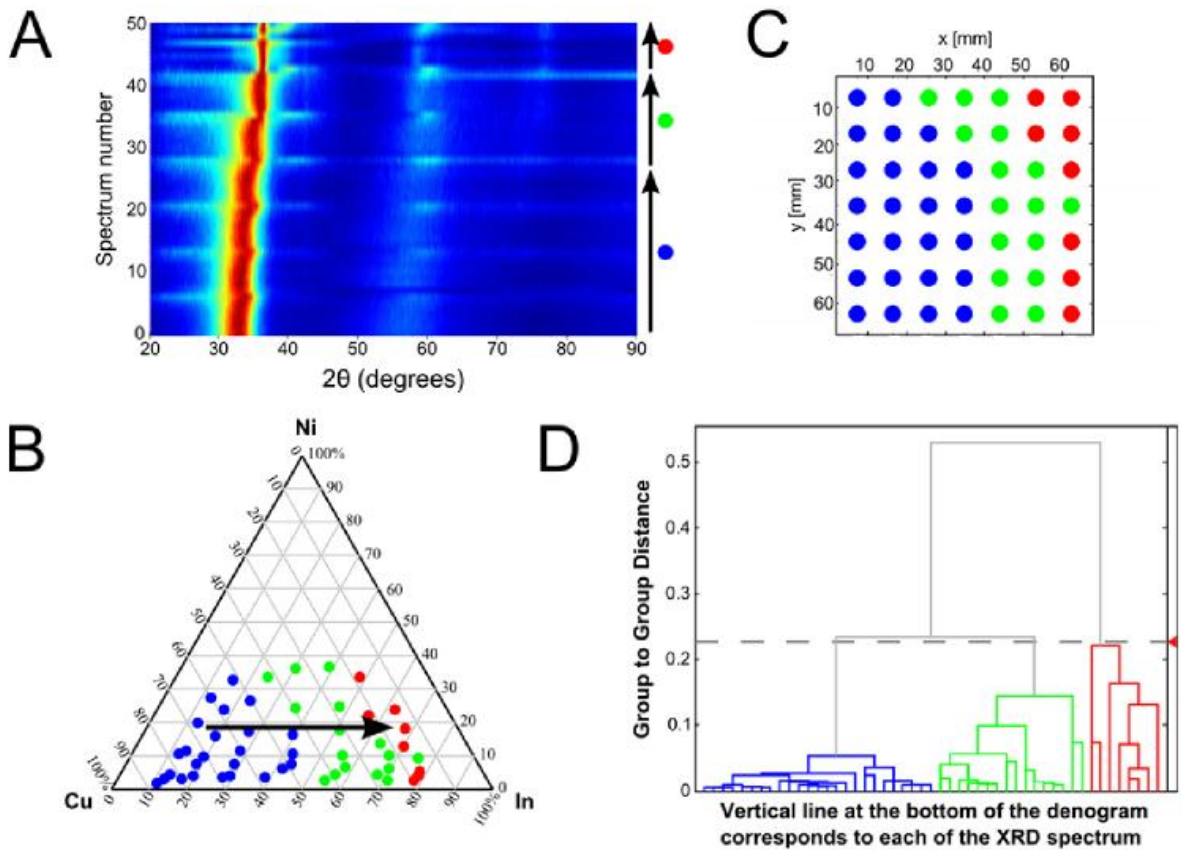

**Figure 20.** a) A map of XRD spectra for all compositions arranged in the following order: blue- group 1, green- group 2 and red- group 3. The transition from one group to anther is shown with arrows; b) a composition phase diagram with an arrow marking the change in thin film crystal structures; c) the grouping of XRD spectra in a rectangular thin film sample; d) a dendogram representing possible groupings of XRD spectra. Each XRD spectrum is represented by a vertical line at the base of the dendogram. The grouping of spectra into larger groups is represented by joining these vertical lines together using horizontal tie lines. The height at which two vertical lines are joined represents the distance between the groups of spectra to be joined, with larger heights corresponding to larger distances. By moving the threshold of group to group distance, the number of groups is adjusted.

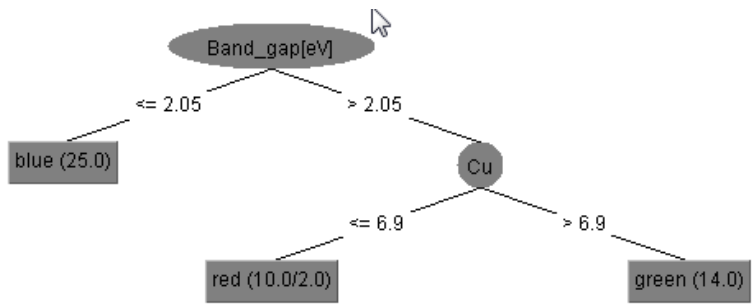

**Figure 21.** Decision tree showing the transition steps between the band gap and three different groups.

The identification of known thin film phases was achieved via comparison of the XRD spectrum for the most representative pattern of each region and a database of known structures to identify known phases (see Figure 22). By using this method, the analysis and classification of 49 spectra was reduced to 3 spectra.

It is important to point out that all three CuO, NiO and $In_2O_3$ thin films that were prepared separately by PLD at 400 °C and oxygen pressure of ~130 mTorr were polycrystalline. However by mixing those materials at 400 °C and oxygen pressure of ~130 mTorr film structure observed for half of the sample area was amorphous (see Figure 22 blue points) which made the phase analysis of the blue group very difficult. For the blue spectra we excluded the formation of $Cu_2In_2O_5$ phase due to limitation in synthesis conditions: i.e. according to Wada[44,42,48], the group could not be synthesized at temperature values below 900°C. The amorphous peaks at ~35° and ~60° could be assigned to two phases: monoclinic CuO (PDF card No: 00-041-0254) and rhombohedral $CuInO_2$ (PDF card No: 01-070-5772), where the last one will be the dominant phase with increase of indium percentage in the group of blue points, which brings the ratio between Cu and In to be around ~1.

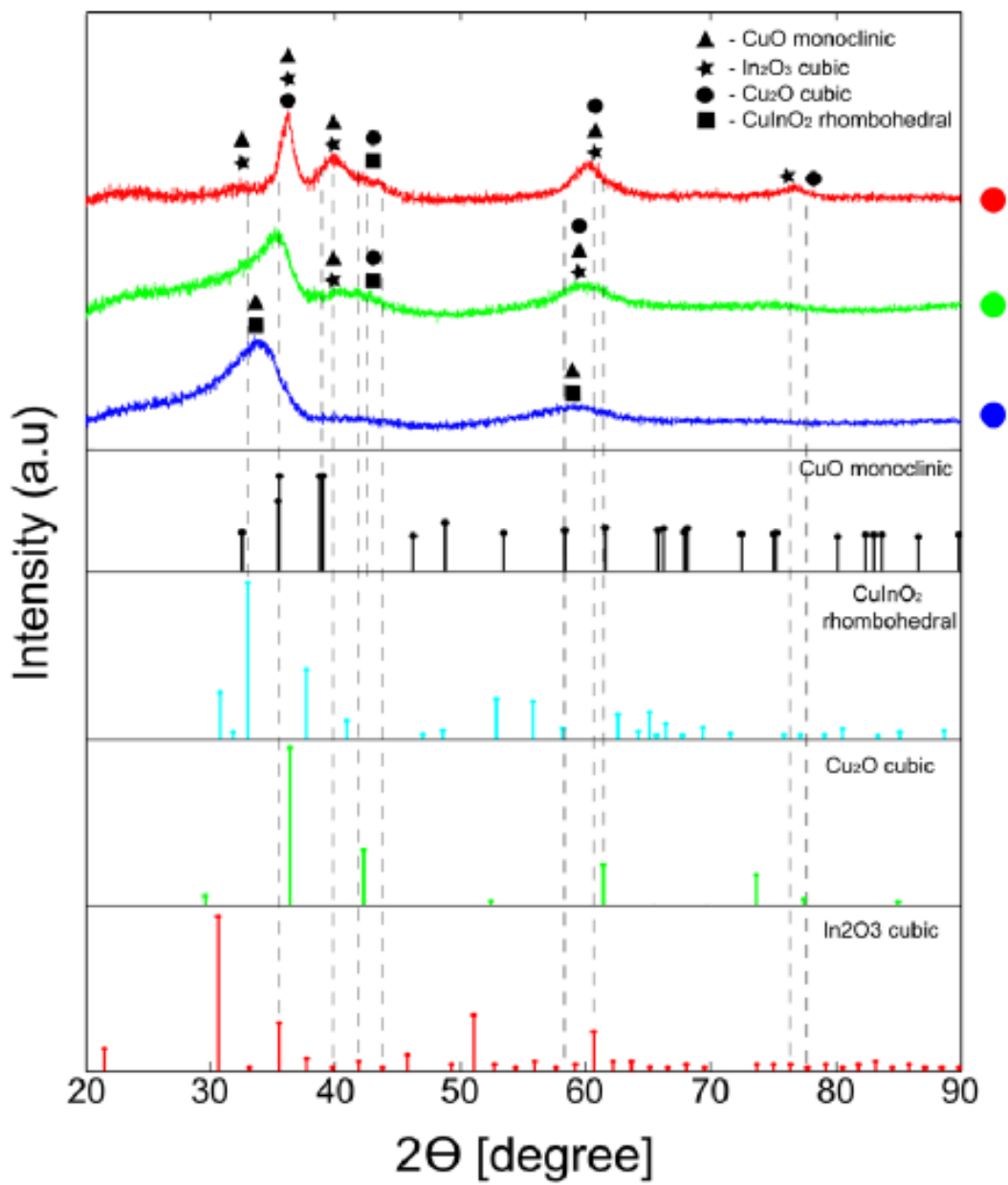

**Figure 22.** The most representative patterns in the blue, green and red groups from Figure 19(a).

The green spectrum corresponds to two possible phases: monoclinic CuO (PDF card No: 00-041-0254), and cubic $In_2O_3$ (PDF card No: 00-006-0416), where the cubic $Cu_2O$ phase, (PDF card No: 01-071-3645), is less probable to dominate due to the low copper concentration. With increase of the indium fraction we observe the shift towards the $In_2O_3$ phase (see Figure 20(a)). The observed result is due to the attribution of the red group to the cubic $In_2O_3$ phase and minor presence of monoclinic CuO phase. We excluded the existence of the rhombohedral $In_2O_3$ phase, because it can hardly be formed under low pressure conditions[49]. Rhombohedral $In_2O_3$ is known as a high pressure and temperature phase.

In all spectra single NiO phase or binary combinations of NiO-CuO or NiO-$In_2O_3$ phases were not detected. This could be due to the small concentration of Ni in the film which may incorporate into the $In_2O_3$ and CuO and be considered a dopant material.

**Optical properties**

The band gap values of the CuO-NiO-$In_2O_3$ system were resolved from the Tauc plot as described earlier in section 5.2. The change in band gap of the CuO-NiO-$In_2O_3$ system from 1.6 eV to 2.65 eV (see Figure 23 (a) and (b)) was mainly determined by the composition and crystal structure of the film. The reported band gap values for monoclinic CuO and cubic $In_2O_3$ are ~1.7eV and ~2.8 eV, which confirm the absence of a pure phase of CuO and $In_2O_3$ (see Figure 23 (b)) in the sample,are in good agreement with the structural analysis. We further mapped out the sample into four regions (see Figure 23 (e)) by using hierarchical clustering analysis of absorptance data. It turns out that the regions obtained from absorptance data of 169 points (see Figure 23 (c) and (d)) are similar to XRD data of 49 points which is shown in Figure 20 (b) and (c). Our observation showed that amorphous film structures, which correspond to blue and red regions in Figure 23 (c), are attributed to two inflection points in the Tauc plot curve (see Figure 24) that correspond to two band gaps: optical (Gap1) and mobility gaps (Gap2).

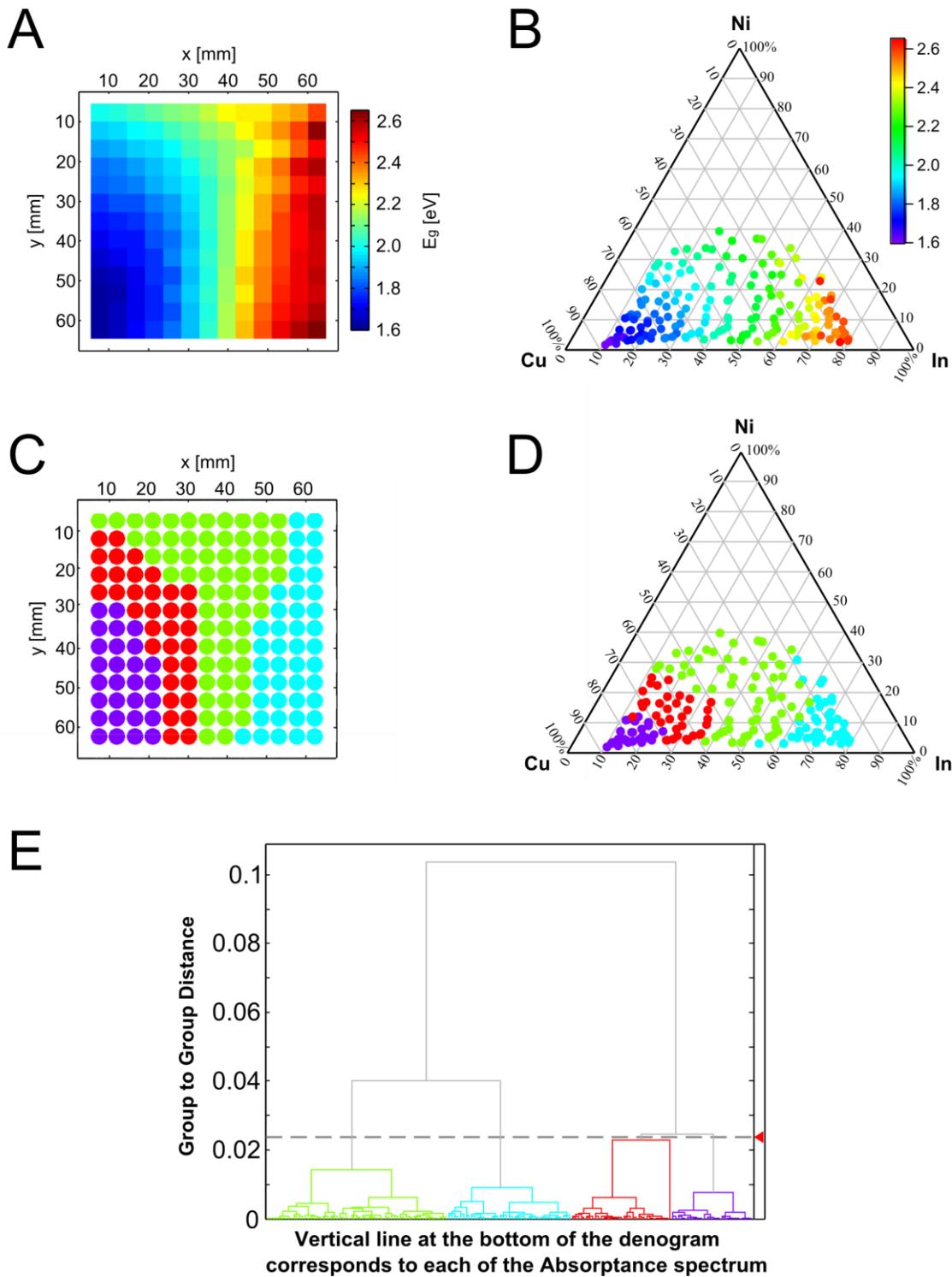

**Figure 23.** a) Plot of direct band gaps as a function of position on the 72 x72 mm sample, showing narrow band gaps in the left bottom corner and wider band gaps in the right side of the sample; b) Ternary diagram showing the composition of CuO-NiO-In$_2$O$_3$ library film, and the measured band gaps extracted from Tauc plot curve; c) the grouping of absorptance spectra in a rectangular thin film sample; d) distribution of clustered absorptance

points in a ternary composition diagram; e) a dendogram representing possible groupings of absorptance spectra.

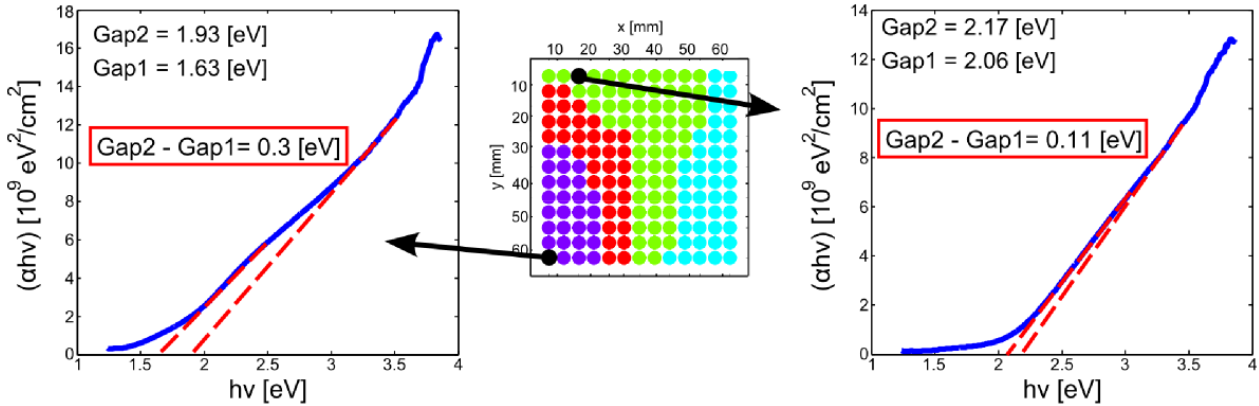

**Figure 24.** Tauc plots at different regions of the sample, where the difference between two band gaps corresponds to the degree of thin film crystallization. The left curve was observed in amorphous copper oxide region, where concentration of Indium metal is from ~10% to 15% and nickel metal is ~4%, leading to the increase of difference between mobility and optical band gaps. The right curve was taken from the region of green points, where concentration of Indium metal is from ~10% to 15% and nickel metal is ~35%, leading to the decrease of difference between mobility and optical band gaps.

In order to explain the above phenomenon we used a Davis-Mott model for amorphous materials[50]. In this band Davis-Mott model, localized states which originate from a lack of long range order extend from the conduction band and valence band to form two ranges, $E_A$ to $E_C$ an $E_V$ to $E_B$, respectively, where $E_C$ and $E_V$ denote the critical energies that separate the extended and the localized states (see Figure 25). The defect states above $E_B$ and under $E_A$ form longer tails. The existence of localized and defect states determine the appearance of Gap1. We observe that an increase in indium content leads to a decrease in short range order (crystallization) so that the position of $E_A$ and $E_B$ moves to $E_C$ and $E_V$, respectively, resulting in an increase of the optical gap as shown in Figure 24.

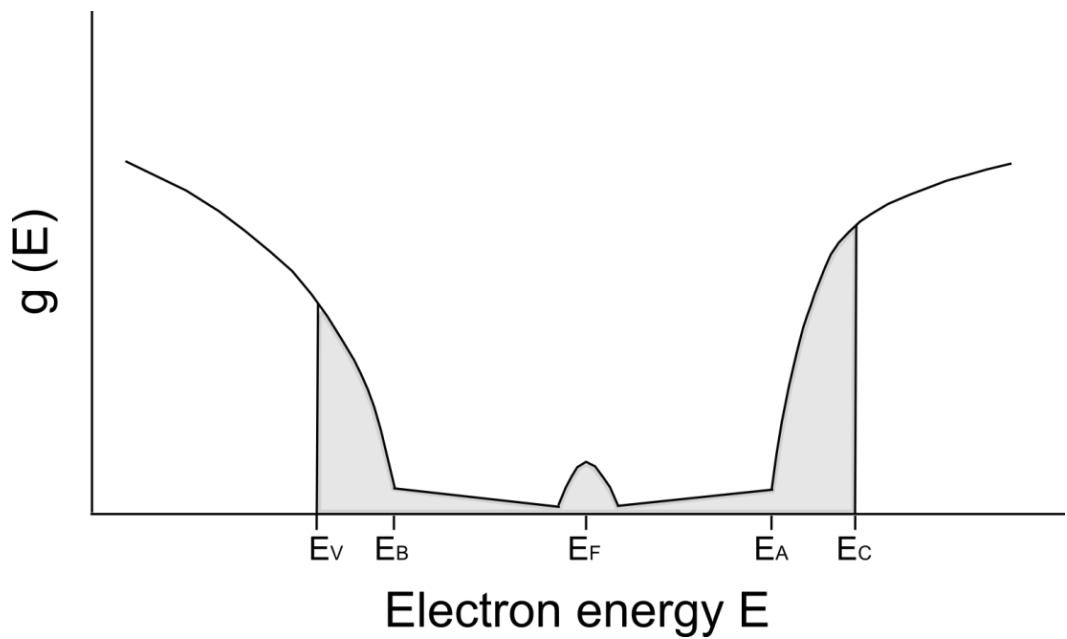

**Figure 25.** Sketch of the Davis-Mott model. Localized states lay in the bands $\Delta E_C$ and $\Delta E_V$ and are due to the lack of long-range order. The tails formed due to the defects in the crystal structure of the film. $E_F$ –is Fermi energy.

Therefore, the increase in mobility gap (Gap2) values accompanied by an increase of the amount of indium is assumed to be a result of the crystallization state of the films. The minimum difference between Gap1 and Gap2 suggests that localized or defect states are reduced by increasing the indium fraction. The dependence of band gaps on crystallization states for some materials was previously reported[51,52].

For points marked in cyan and green, only one band gap was observed, indicating that the thin film has a polycrystalline structure where the Davis-Mott model is not applicable. For complex systems made of more than one material the inflection in the Tauc plot curve could appear due to the different material phases. However, we further show (section 5.4) that the presence of both band gap values is due to the amorphous nature of the film in which only a mobility gap (Gap 2) contributes to photovoltaic activity of the solar cells.

**Electrical analysis**

In addition to the optical analysis, the effects of temperature and light on electrical properties of the film were studied. The resistances were only detected in the amorphous part of the film corresponding to different material compositions (see Figure 26 (a)). We observed an increase in resistance from 0.2 G$\Omega$ to 1.2 G$\Omega$ that correlated to an increase in indium (In) concentration from 10% to 50%. Since $In_2O_3$ is known as a highly resistive material, the resistance values higher than 2G$\Omega$, where the indium (In) concentration exceeds 50%, were not detected, and that indicated the dominance of the cubic $In_2O_3$ phase.

Conductivity, an intrinsic property of the film, was calculated by using equations (1) and (2). High conductivity values of the sample were obtained for material concentrations of Cu (from 90% to 60%), In (from 10% to 40 %) and Ni (from 2% to 35%) (see Figure 26 (b)). Interestingly, when the ratio between resistances was measured with and without light (see Figure 26 (c)), it was observed that an increase of Ni metal atoms in the film from 10% to 40% led to a high resistance ratio, which is attributed to the large amount of electron/hole pairs that were generated at this region. We conclude from Figure 26 (c) that Ni atoms play a significant role as a doping element in the CuO-NiO-$In_2O_3$ system, and exhibit high electron/hole generation due to the introduction of light. Moreover, the ratio between resistances calculated by using equation (11) provides information on material composition which gives high photovoltaic performance once the film is introduced in a solar cell device.

$$\sigma_{ratio} = \left(\frac{R_{dark} - R_{light}}{R_{dark}}\right) * 100 \qquad (11)$$

Where $\sigma_{ratio}$ is the ratio between resistances with and without light, $R_{dark}$ − is the resistance measured in dark, and $R_{light}$ - is the resistance measured with light.

In order to understand the electrical conduction mechanisms of the $CuO$-$NiO$-$In_2O_3$ system, activation energy was derived from the temperature dependent conductivity plot from 39.6 °C to 247 °C in air (see Figure 26 (d-f)). The activation energies $Ea_1$ and $Ea_2$ were measured at different temperature ranges and resolved for different semiconductor conduction mechanisms (see Figure 26 (d)).

For the amorphous film structure described in Figures 20 (b-c) and Figure 22, activation energies of the sample at a temperature range of 39.6 °C to 89 °C varied from 0.05 eV to 0.2 eV. These values are lower than the activation energy of the CuO thin film prepared by pulsed laser deposition at 400 °C (0.28 eV). Jeong and Choi[53] reported that conduction of $Cu_{1-y}O$ is due to the hopping of the charge carrier with activation energy of 0.1 eV that suggests the activation was due to electrical mobility. The trend of the activation energy at lower temperatures is consistent with the samples' conductivity, which suggests that smaller activation energy results in higher conductivities of the sample and conduction at temperatures in the range of 36.9 °C to 89 °C is due to the hopping of electrons (see Figure 26 (b) and (f)).

For the high temperature values of 89 °C - 247 °C we observed an increase in activation energies with the increase inconcentration of indium (In) atoms (see Figure 26 (e)). According to the model of density of states in amorphous semiconductors, the activation energy for temperature values 89 °C - 247 °C suggests that the excitation of electrons to $E_C$ or holes to $E_V$, produces currents, as given by equation (6).

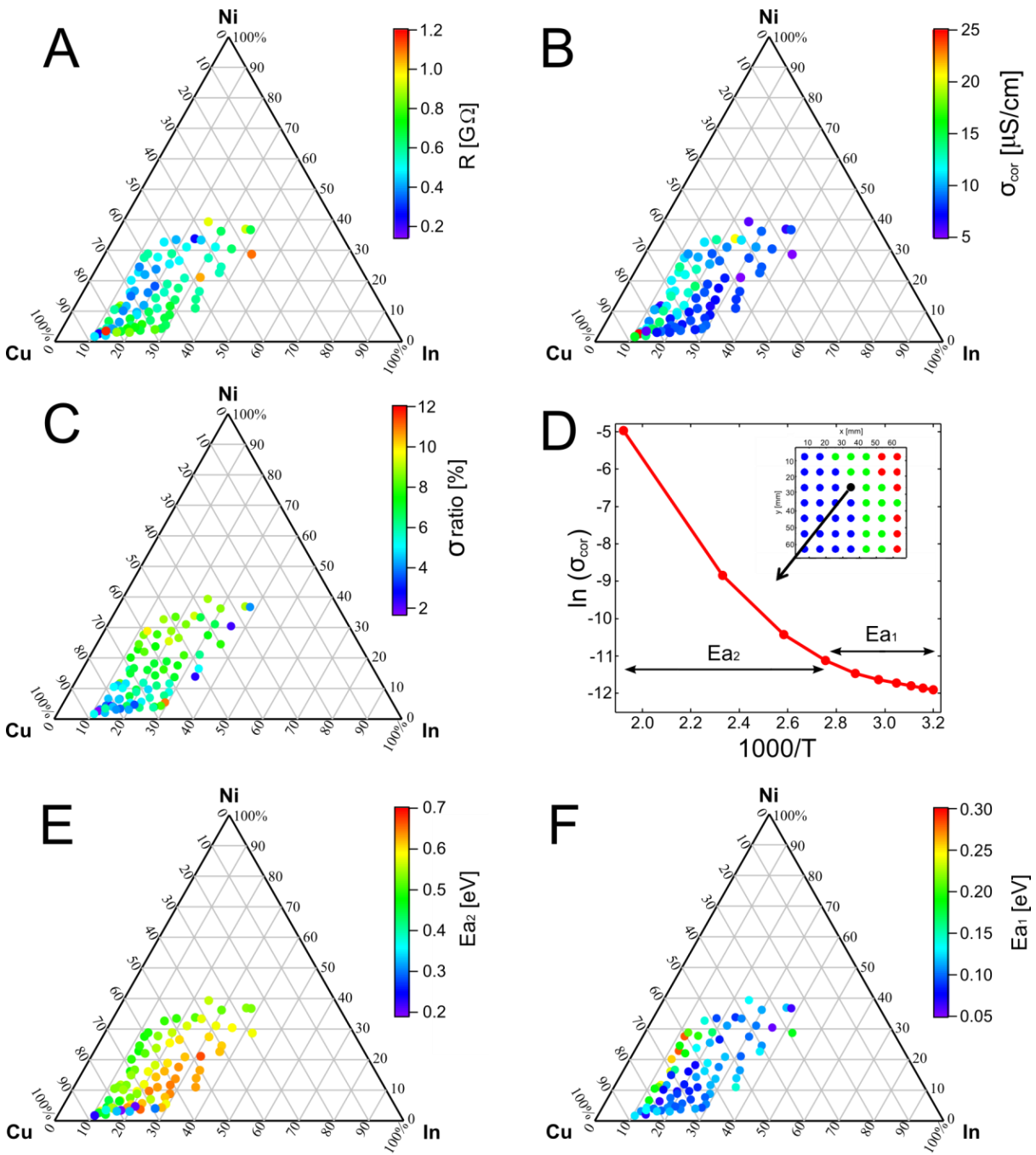

**Figure 26.** a) Resistance map of CuO-NiO-In₂O₃ system; b) corrected conductivity composition map at 39.6 °C; c) Ratio between resistivity measured with and without light; d) Arrhenius plot of single point corresponding to the XRD map; e) Activation energy calculated at temperature ranges of 89 °C - 247 °C; f) Activation energy calculated at temperature ranges of 39.6 °C – 89 °C.

# 5.4 Photovoltaic activity of CuO-NiO-In$_2$O$_3$ solar cell libraries

In order to evaluate the charge separation and voltage generation in a CuO-NiO-In$_2$O$_3$ absorber film, asymmetric junctions of TiO$_2$ / CuO-NiO-In$_2$O$_3$ with Au and Ag metal back contacts were produced as libraries. A complete analysis of the PV parameters *Jsc*, *Voc*, *FF*, *Pmax*, *Rs* and *Rsh* for libraries containing Ag and Au metal back contacts is shown in Figures 27 and 29. The *Jsc* ternary composition map (see Figure 27(a)) shows low current densities in the presence of high fractions of (Cu) metal which increases with the increase of (Ni) and (In) contents. The *Voc* composition map has a similar behavior to the *Jsc* composition map. At regions with high (Cu) concentration, low values for *Jsc* and *Voc* were obtained, suggesting that monoclinic CuO and rhombohedral CuInO$_2$ phases, previously described as dominant, play a minor role in the photovoltaic activity of the library cells. Therefore it appears that the Ni fraction being used as a doping material is the factor that affects the *Jsc*, *Voc*, *Pmax* and *FF*. The majority of the *Voc* ternary composition map exhibits values from 450 mV to 550mV (see Figure 27 (b)), while the high fill factor (*FF*) values of 38% - 45% in the composition map were achieved through adding Ni of 20% to 40%. The maximum power density, *Pmax*, indicates the region corresponding to the green XRD group which is presented in Figure 22, and shows that the proposed dominant polycrystalline CuO and In$_2$O$_3$ phases is where Pmax is strongly dependent on nickel concentration.

The shunt, *Rsh*, and series, *Rs*, resistances of a ternary composition diagram were derived from I-V curves under illumination and are shown in Figure 27 (e-f). The series resistance increases with a decrease of the Ni fraction from 20 K$\Omega$ cm$^2$ to 700 K$\Omega$ cm$^2$ at the edge of the ternary diagram between the Cu and In components (see Figure 27e). At a high concentration of indium (In), the series resistance *Rs* changes significantly from 200 K$\Omega$ cm$^2$ to 0.5 K$\Omega$ $cm^2$, compared to the rest of the sample library. However, the series resistance doesn't reach the highest fill factor (*FF*) of the sample library[54] due to the high indium fraction corresponding to the cubic $In_2O_3$ phase which is known as a non-photoactive material due to its wide band gap of 3.2 eV.

In order to achieve good solar cell operation, however, the shunt resistances should be larger than the Rs. For the $TiO_2$ / $CuO$-$NiO$-$In_2O_3$ hetero junction library, values between 0.5 and 3 $M\Omega\ cm^2$ were observed under illumination.

The Internal Quantum Efficiency (*IQE*)[25] which is shown in Figure 27 (g) provides information on the charge separation and collection efficiencies of the hetero junction library devices. Maximal *IQE* values of 0.3% reveal the poor charge transport mechanisms across the junction interface between $TiO_2$ and $CuO$-$NiO$-$In_2O_3$ thin films. However, relatively high *IQE* values in the ternary composition diagram were obtained where the nickel (Ni) fraction is high. The *IQE* is the ratio between measured (*Jsc*) and calculated (*Jcalc*) short circuit current densities which are shown in Figure 27 (a and h).

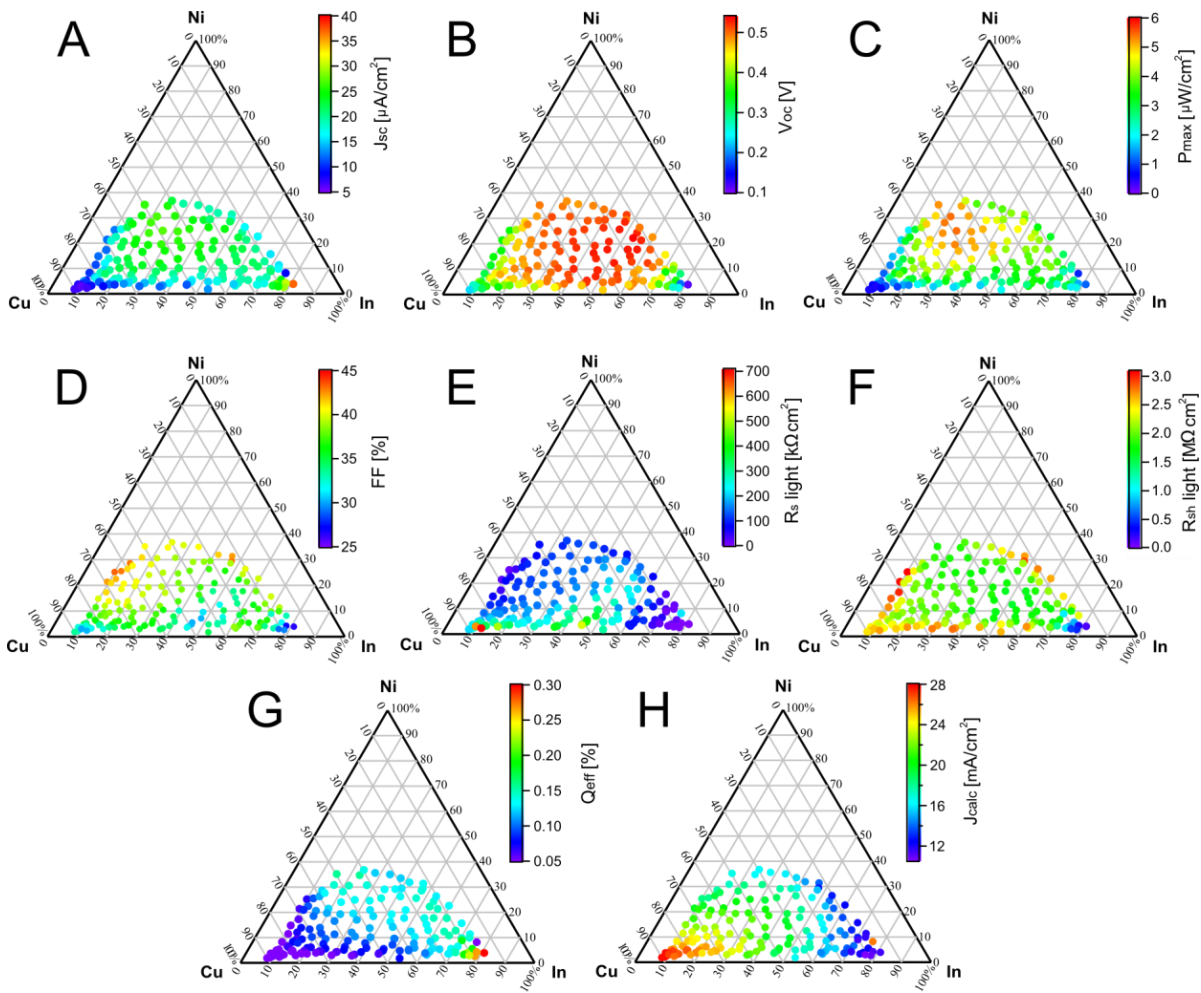

**Figure 27.** Ternary composition maps of the photovoltaic parameters for Ag metal back contact: a) *Jsc*, b) *Voc*, c) *Pmax*, d) *FF*, e) *Rs* measured under light, f) *Rsh* measured under light, g) Internal quantum efficiency (*IQE*), h) maximum theoretical short circuit photocurrent density (*Jcalc*).

To confirm the significant role of the nickel fraction in PV activity of a $TiO_2$ / CuO-NiO-$In_2O_3$ hetero junction solar cell, we fabricated a $TiO_2$ / CuO-NiO hetero junction solar cells library. The $TiO_2$ / CuO-NiO hetero junction library revealed only 4 photovoltaic points with low *Jsc*, *Voc*, *FF* and *Pmax* values.

For the $TiO_2$ /CuO-NiO-$In_2O_3$ hetero junction library with Au metal back contacts (see Figure 28) the *Jsc*, *Voc*, *FF*, *Pmax*, *Rs*, and *Rsh* parameters showed the same behavior as for Ag contacts. The difference between Ag and Au contacts are presented in Figure 29 as a ratio between each of the photovoltaic parameters. The ratios of *Jsc*, *Voc*, *FF*, and *Pmax* for the majority of the sample library are above 1, indicating the improvement of PV parameters using an Ag contact.

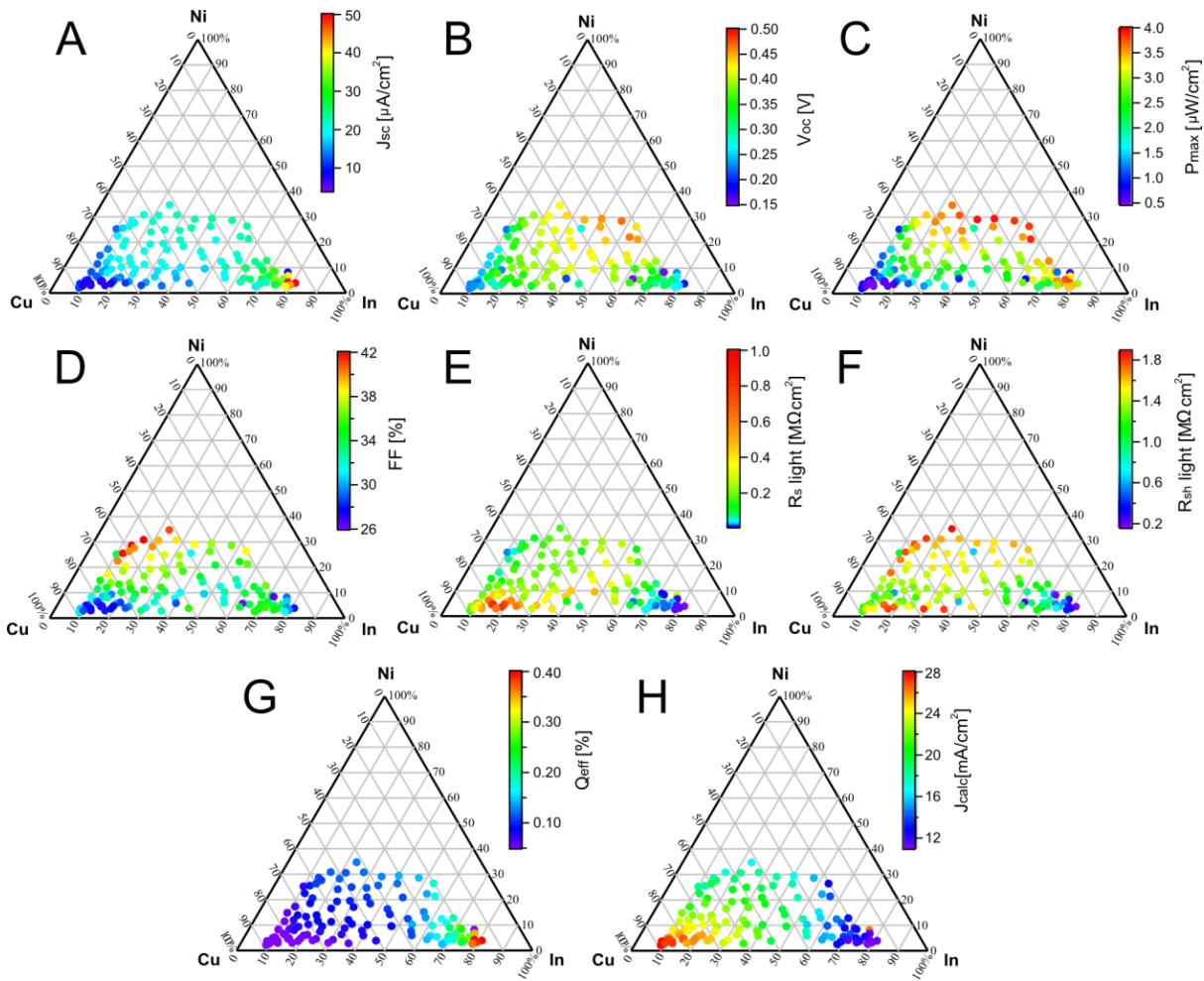

**Figure 28.** Ternary composition maps of the photovoltaic parameters for Au metal back contact: a) *Jsc*, b) *Voc*, c) *Pmax*, d) *FF*, e) *Rs* measured under light, f) *Rsh* measured under light, g) Internal quantum efficiency (*IQE*), h) maximum theoretical short circuit photocurrent density (*Jcalc*).

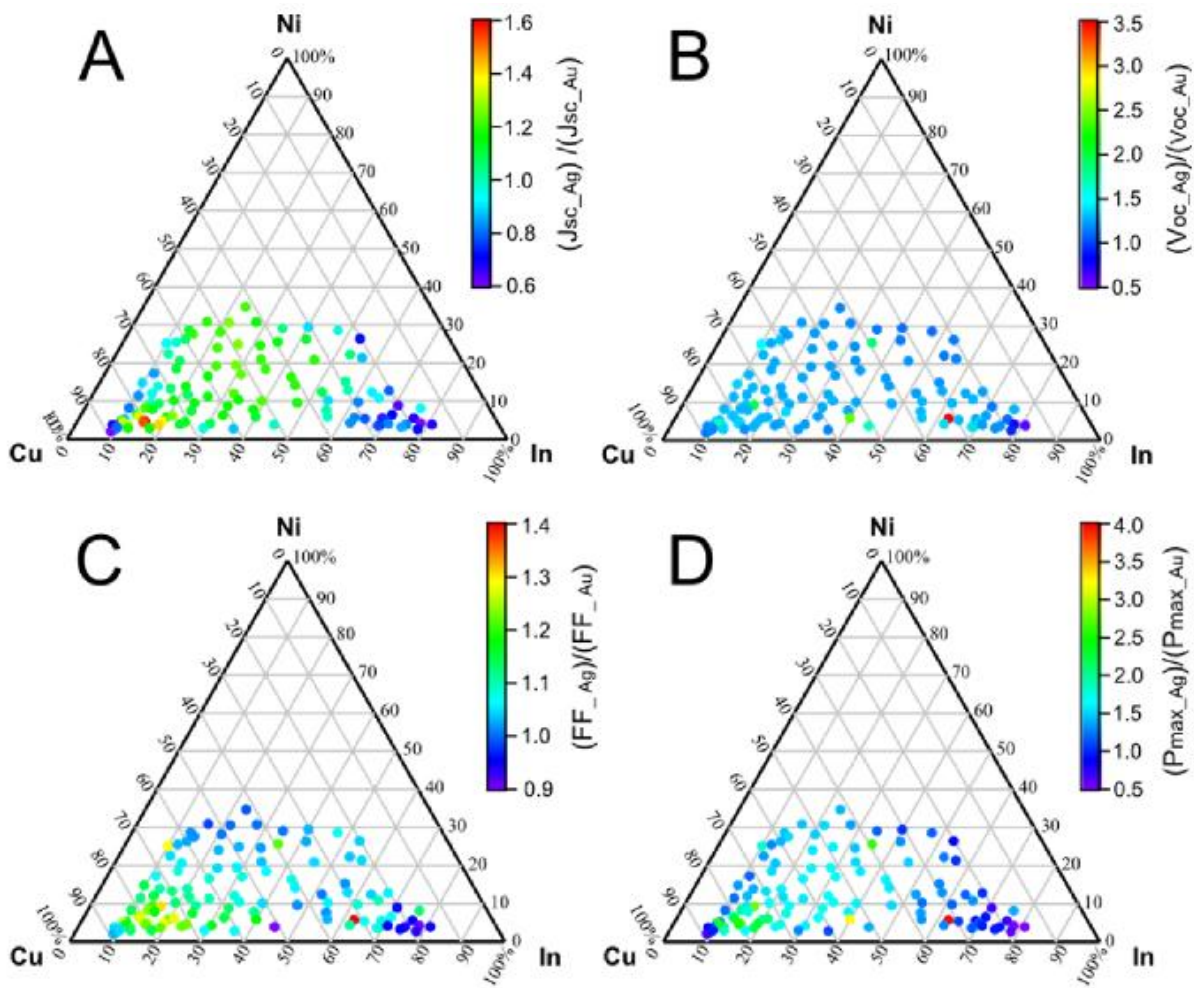

**Figure 29.** The ternary composition maps ratios between Ag and Au contacts of: a) *Jsc* ; b) *Voc*; c) *FF*; d) *Pmax*

At the front part of a solar cell, the TiO$_2$ layer acts as an electron transport layer as well as a window to allow light to pass through to the CuO-NiO-In$_2$O$_3$ absorber layer. The NiO at the back acts as a hole transport layer which improves solar cell library performance for Ag and Au metal back contacts by providing a better charge separation (see Figures 30 and 31).

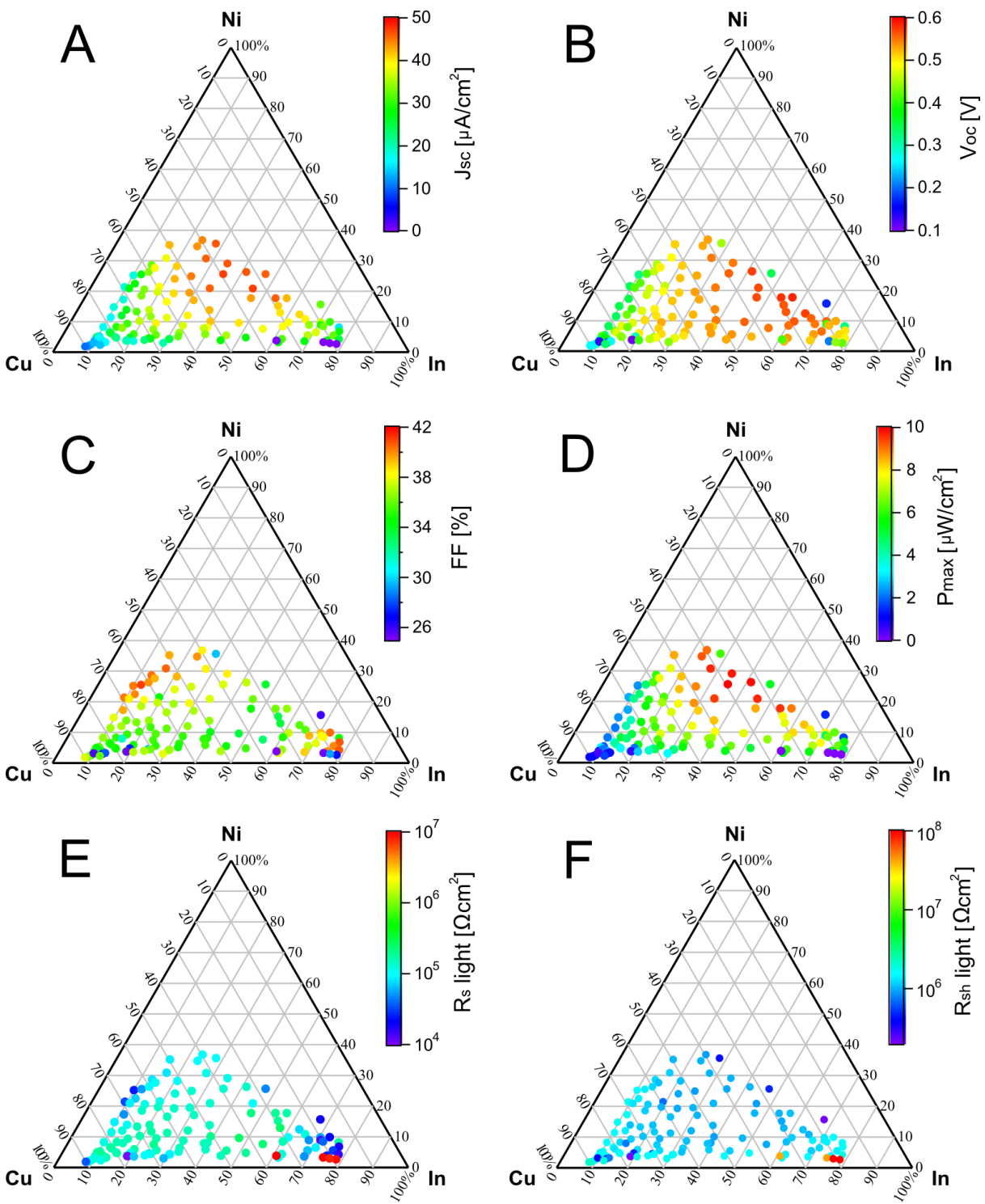

**Figure 30.** Ternary composition maps of the photovoltaic parameters for **Ag** metal back contacts deposited on **NiO** hole conductive thin film: a) *Jsc*, b) *Voc*, c) *Pmax*, d) *FF*, e) *Rs* measured under light, f) *Rsh* measured under light.

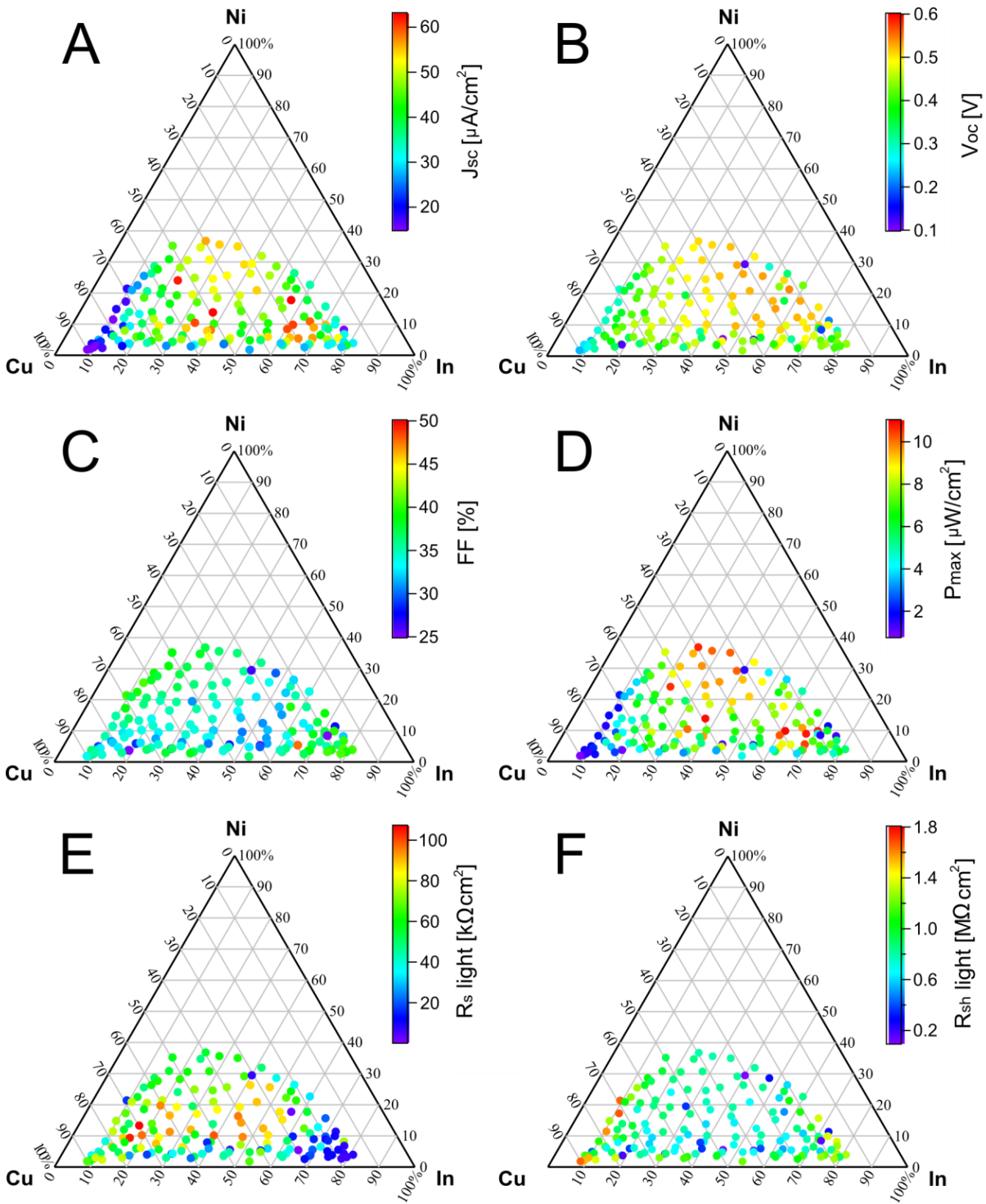

**Figure 31.** Ternary composition maps of the photovoltaic parameters for **Au** metal back contact deposited on **NiO** hole conductive thin film: a) *Jsc*, b) *Voc*, c) *Pmax*, d) *FF*, e) *Rs* measured under light, f) *Rsh* measured under light.

# Chapter VI: Summary and Conclusions

In this thesis I have demonstrated a combinatorial material science approach for the development and rapid characterization of new and known metal oxide thin films for all oxide photovoltaics. This approach deals with large datasets that contain a significant amount of information which are performed in a relatively short time. Such techniques are generally achieved through the use of serial automation or by conducting several measurements simultaneously. Therefore, I discussed the development of a materials' electrical characterization tool called "Electrical conductivity scanning system". Using this system I performed the mapping of the conductivity and activation energy of large sample areas, of about 72 x 72 mm$^2$. I identified the electrical properties of both resistive metal oxide thin film samples with high precision and accuracy. The scanning system enabled me to gain insight into the transport mechanisms of single CuO, NiO and ternary CuO-NiO-In$_2$O$_3$ thin film materials, and to use those insights to make smart choices in the design of multilayer thin film all oxide photovoltaic cells.

In the second part of the my thesis the nickel oxide thin film with the most desirable electrical, optical and structural properties for use in all oxide solar cells was discussed. The correlation between electrical, optical and structural properties of pulsed laser deposited NiO thin film was observed.

In the third part the ternary CuO-NiO-In$_2$O$_3$ metal oxide system was introduced for use in hetero junction all oxide solar cells' structure as an absorber material. Using agglomerative hierarchical clustering, the structural, optical, and electrical characterizations were greatly accelerated and allowed efficient material analysis. The analysis revealed that Ni atoms play a significant role as a doping element in the CuO-NiO-In$_2$O$_3$ system, which exhibits high electron/hole generation due to the introduction of light.

The CuO-NiO-In$_2$O$_3$ system was found to be a new promising absorber material providing a relatively high open voltage circuit (*Voc*) of 0.55 eV in a TiO$_2$ / (CuO-NiO-In$_2$O$_3$) hetero junction device library. Photocurrent densities up to 40 µA cm$^{-2}$ were observed. These photocurrent density values were found to be far below the most efficient reported values for n-Ga$_2$O$_3$/p-Cu$_2$O hetero junction all oxide solar cells. The low values for Internal Quantum Efficiency (*IQE*) provide information on the poor charge transport mechanisms across the TiO$_2$ / CuO-NiO-In$_2$O$_3$ junction interface. Furthermore, my observation showed the improvement in *Voc* and *Jsc* values once an NiO hole transport thin film was introduced into the device structure. Therefore, TiO$_2$ / CuO-NiO-In$_2$O$_3$ / NiO hetero junction all oxide PV cells are a promising device structure for photovoltaics, but the photovoltage and especially the photocurrent still leave much room for improvement. The successful application of combinatorial material science into thin film photovoltaics was presented in this thesis.

# Chapter VIII: Bibliography

## תקציר

גישה קומבינטורית לחקר התקנים ו חומרים עשויים מתחממוצות מתכתיות מבוססת על הסינתזה של מאות חומרים הקשורים בניסוי אחד. הגישה מצריכה פיתוח כלי מדידה וטכניקות חדשות, על מנת לאפיין במהירות ספריות של חומרים ולנתח את הנתונים המתקבלים. מחקר המוצג כאן נועד לתרום למאמץ לעמוד בדרישה זו, ובכך לקדם את קצב חקר חומרים. במקרים רבים הגדרות פוטו וולטאיות מותאמות היטב לשיטות תפוקה גבוהה, המאפשרות ניתוח כמותי ישיר של תכונות פיסיקליות אשר את יישומם הפגנתי בחלק הראשון של התזה.

חלק הראשון של התזה דן בבנייה ו פיתוח כלי מדידה לאפיון חומרים שנקרא "מערכת סריקת מוליכות חשמלית", ששימשה לקביעת התנגדות סגולית, מוליכות, אנרגיית הפעלה עבור ספריות של תחממוצות מתכתיות מרוכבות.

חלק שני של התזה דן בפתרון המרחבי של תכונות אופטיות וחשמליות עבור שכבת תחממוצת ניקל לא הומוגנית בעובי שהוכנה על ידי שיטה הנקראת (PLD). בנוסף לאפיון מפורט של תכונות מבניות, חשמליות, ו אופטיות עבור שכבת תחממוצת ניקל אני מציג את קשר בין אנרגיית הפעלה לגורם המרקם. התוצאות מעידות על כך ששכבה דקה של תחממוצת ניקל יכולה לשמש כמגע סלקטיבי בתאי שמש מרובי שכבות המבוססים על תחממוצות מתכתיות. בנוסף שכבה זו יכולה לשפר את ביצועים פוטו וולטאיים כתוצאה מיכולת העברה סלקטיבית של מטען חופשי ו יציבותה בסביבה חיצונית.

בס''ד

# אוניברסיטת בר אילן

## גישה קומבינטורית לפיתוח של תחמוצות מתכתיות חדשות

## עבור תאים פוטו - וולטאים

קלימנטי שימנוביץ'

עבודה זו מוגשת כחלק מהדרישות לשם קבלת תואר מוסמך במחלקה

לכימיה של אוניברסיטת בר-אילן.

רמת-גן, ישראל                                      תשע"ד